\documentclass[sn-mathphys-num,iicol]{sn-jnl}%
\usepackage{graphicx}%
\usepackage{multirow}%
\usepackage{amsmath,amssymb,amsfonts}%
\usepackage{amsthm}%
\usepackage{mathrsfs}%
\usepackage[title]{appendix}%
\usepackage{xcolor}%
\usepackage{textcomp}%
\usepackage{manyfoot}%
\usepackage{booktabs}%
\usepackage{algorithm}%
\usepackage{algorithmicx}%
\usepackage{algpseudocode}%
\usepackage{listings}%
\usepackage{siunitx}
\usepackage{subcaption}
\usepackage{tikz}
\usepackage{pgfplots}
\usepackage{mathabx}
\usepackage{mhchem}        

\usetikzlibrary{tikzmark}
\usetikzlibrary{positioning}
\usetikzlibrary{decorations.pathreplacing}
\usetikzlibrary{shapes.geometric}
\usepgfplotslibrary{fillbetween}
\usetikzlibrary{patterns} 

\theoremstyle{thmstyleone}%
\theoremstyle{thmstyletwo}%

\theoremstyle{thmstylethree}%

\DeclareMathOperator{\Tw}{Tw}
\DeclareMathOperator{\E}{E}

\begin{document}

\title{First-principle event reconstruction by time-charge readouts for TAO}

\author[1,2,3]{\fnm{Xuewei} \sur{Liu}}

\author[1,2,3]{\fnm{Wei} \sur{Dou}}

\author*[1,2,3,4]{\fnm{Benda} \sur{Xu}}\email{orv@tsinghua.edu.cn}

\author[5]{\fnm{Hanwen} \sur{Wang}}

\author[5]{\fnm{Guofu} \sur{Cao}}

\affil[1]{\orgdiv{Department of Engineering Physics}, \orgname{Tsinghua
    University}, \orgaddress{\city{Beijing}, \country{China}}}

\affil[2]{\orgdiv{Center for High Energy Physics}, \orgname{Tsinghua
    University}, \orgaddress{\city{Beijing}, \country{China}}}

\affil[3]{\orgdiv{Key Laboratory of Particle \& Radiation Imaging (Tsinghua
    University)}, \orgname{Ministry of Education}, \orgaddress{\country{China}}}

\affil[4]{\orgdiv{Kavli Institute for the Physics and
    Mathematics of the Universe, UTIAS}, \orgname{the University of Tokyo},
  \orgaddress{\city{Tokyo}, \country{Japan}}}

\affil[5]{\orgdiv{Institute of High Energy Physics}, \orgname{Chinese Academy of Sciences},
  \orgaddress{\city{Beijing}, \country{China}}}

\abstract{
  The Taishan Antineutrino Observatory (TAO)
is a liquid-scintillator satellite experiment of the Jiangmen Underground
Neutrino Observatory (JUNO)
to measure the reference reactor neutrino spectrum
with unprecedented energy resolution.
We use inhomogeneous Poisson process and
Tweedie generalized linear model (GLM)
to characterize the detector response and the charge distribution
of a SiPM.
We develop a pure
probabilistic model for time and charge
of SiPMs from first principles to reconstruct point-like events in the TAO central detector.
Thanks to our precise model and the high photo-coverage and quantum efficiency
of the SiPM tiles at TAO, we achieve
vertex position resolution better than \SI{20}{mm},
energy resolution of about \SI{2}{\percent} at \SI{1}{MeV} and \SI{<0.5}{\percent} non-uniformity,
marking the world's best performance of liquid scintillator detectors.
With such resolution, we perceive \si{MeV} events to exhibit track effects.
It opens up an exciting possibility of computed tracking calorimeter
for unsegmented liquid scintillator detector like TAO.
Our methodology is applicable to other experiments
that utilize PMTs for time and charge readouts.


}

\maketitle

\section{Introduction}

The Taishan Antineutrino Observatory (TAO) is a satellite experiment
of the Jiangmen Underground Neutrino Observatory
(JUNO)~\cite{xu_calibration_2022}.
Using \SI{2.8}{tons} Gadolinium-doped
Liquid Scintillator (GdLS) and 4024 Silicon Photomultiplier (SiPM)
tiles, TAO will measure the neutrino energy spectrum with unprecedented precision from a
reactor core of the Taishan Nuclear Power Plant \SI{44}{m} away.
The neutrino energy spectra predicted from recent computations~\cite{PhysRevC.84.024617,PhysRevC.83.054615}
disagree with the ones measured
by the previous reactor neutrino experiments such as Daya Bay~\cite{PhysRevLett.116.061801},
Double Chooz~\cite{DoubleChooz:2019qbj}, RENO~\cite{PhysRevLett.121.201801},
NEOS~\cite{PhysRevLett.118.121802}, STEREO~\cite{STEREO:2022nzk}.
The inconsistency is believed to have its roots in lack of complete information on decay and fission
yields from the nuclear database~\cite{PhysRevLett.123.111801}.
To determine the neutrino mass ordering,
JUNO demands TAO for model-independent reference spectra~\cite{JUNO:2021vlw}.
The TAO collaboration plans to benchmark the nuclear database with fine
structures in the antineutrino spectra~\cite{PhysRevC.98.014323}.
Thus, we require the position resolution to be better than
\SI{5}{cm}, energy resolution to be \SI{\sim 2}{\percent} at
\SI{1}{MeV} and energy non-uniformity to be contained within
\SI{0.5}{\percent} after \emph{event reconstruction}~\cite{xu_calibration_2022,JUNO_2020ijm}.

In many large liquid scintillation and water Cherenkov
detectors, the arrival time of the first photo-electron~(PE)
and the total
integrated charge in a chunk of PMT/SiPM readout
waveforms are input to the event reconstruction stage of data reduction.
The time distribution of the first PE is long known to be affected by PE
pile-up, where multiple PEs arriving in rapid succession cannot be
distinguished, thereby distorting the time distribution~\cite{RANUCCI1995389,GALBIATI2006700}.
KamLAND~\cite{batygov_combined_2006}
uses a time-only vertex fitter with heuristic corrections.
Borexino~\cite{PhysRevD.89.112007} and
Super-Kamiokande~\cite{10.1093/ptep/ptz015} construct
several empirical first-PE time probability density functions~(PDF) from both calibration and Monte Carlo
conditioned on charges.
Z.~Li~et~al~\cite{ziyuan2021} derive a
rigorous time dependence on the PE counts for JUNO. However, the counts are
inaccurately estimated from rounding charges to integers. G.~Huang~et~al.~\cite{guihong2023}
improve upon it by relying on both the PE count and expectation of it.
But the time-charge-combined likelihood is an oversimplified direct product
assuming independence of the two components.
Such approximations introduce inherent bias needing to be \textit{ad-hocly} corrected \textit{a posteriori}
in form of \emph{correction maps}.
Z.~Qian~et~al.~\cite{QIAN2021165527} and Gavrikov~et~al.~\cite{Gavrikov:2022kok} discuss the application of
several convenient and flexible end-to-end machine learning models, though the performance of
which depends on high-fidelity Monte Carlo, selection of aggregated
features and optimal hyperparameters.
It is challenging to quantitatively assess the
degree to which the algorithmic framework itself contributes to the observed
non-uniformity in the reconstruction results.

To address those difficulties, a fundamental model derived from first
principles is necessary, especially for an experiment like TAO with
unprecedented energy and vertex resolution.  We use Tweedie
generalized linear model~(GLM) to describe the probabilistic relation
of PE count and charge.  Upon it, we derive an exact joint time-charge
PDF from the original light curve.  The resulting reconstruction
algorithm is free from correction maps and hyperparameters.  It is
transparent in that the non-uniformity of the reconstructed energy is
entirely determined by the detector calibration.
Sec.~\ref{sec:detector_response} discusses the definition and
implementation of the detector response for a point-like event in the TAO central
detector. Sec.~\ref{sec:TweedieGLM} derives the exact time-charge likelihood
from the Tweedie distribution. Sec.~\ref{sec:dataset} introduces the dataset
before evaluating the bias and resolution of the reconstructed position
and energy.  Sec.~\ref{sec:discussion} discusses the limitations of
our approach and future improvements.  Finally, we conclude in Sec.~\ref{sec:conclusion}.


\section{Optical detector model}
\label{sec:detector_response}

Fig.~\ref{fig:tao_cd} shows the schematic of TAO central detector~(CD).
A spherical acrylic vessel with an inner diameter of
\SI{1.8}{m} is filled with \SI{\sim2.8}{tons} GdLS.
The GdLS is composed of Linear Alkylbenzene~(LAB) as the solvent, supplemented
with \SI{2}{g/L} of 2,5-Diphenyloxazole~(PPO) as the fluor and \SI{1}{mg/L} of
p-bis-(o-Methylstyryl)-benzene~(bis-MSB) as the wavelength shifter. The mixture
is doped with gadolinium at a mass fraction of \SI{0.1}{\%}.
The fiducial volume expands to the radius of \SI{0.65}{m},
\SI{0.25}{m} away from the boundary of acrylic vessel.
A total number of 4024 \(50.7 \times \SI{50.7}{\mm^2}\)
SiPM tiles with around fifty percent photon detection efficiency are installed
on the inner surface of copper shell supporting the
acrylic vessel. The copper shell is immersed in a linear alkylbenzene~(LAB) buffer
inside a cylindrical stainless-steel tank.  We focus on the TAO CD and refer
other sub-systems
to H.~Xu~et~al.~\cite{xu_calibration_2022} and Abusleme~et~al.~(JUNO collaboration)~\cite{JUNO_2020ijm}.
TAO detector is under
construction and we deploy Monte Carlo~(MC) simulation to train the
detector response
and evaluate the reconstruction algorithm~(Sec.~\ref{sec:dataset}).

\begin{figure}[htbp]
  \centering
  \includegraphics[width=0.4\textwidth]{figures/Fig1_tao_cd.pdf}
  \caption{Schematic of the TAO central detector.}
  \label{fig:tao_cd}
\end{figure}


The detector response is defined as a map from a point-like
event to the time-charge distributions on SiPM tiles.  We
divide it into two stages.  In this section, the first stage of the response
function is optical.  It maps an event
to PE times for a SiPM, which is properly described by an inhomogeneous Poisson point
process. We utilize the approach developed by W.~Dou~et~al.~\cite{Dou2022ReconstructionOP}
to characterize the optical properties of the
detector including the GdLS time profile
and photon transmission.  The second stage is
the electronics.  It maps
the count and times of the PEs in a SiPM to the first-PE time and the total charge,
modeling the SiPM and analog-to-digital system.
We shall discuss it in Sec.~\ref{sec:TweedieGLM}.

\subsection{Poisson point process}
\label{subsec:res_fun}

Consider the response function of a point-like event \(\delta
\left(\vec{r}, E\right)\) on \(j\)th SiPM, where \(\vec{r}\) and \(E\)
are the vertex and energy of the event.  The occurrence of
PE on \(j\)th SiPM follows an inhomogeneous Poisson
process with intensity function \(R_j(t;\vec{r}, E)\)~\cite{DONATI1969251}.
The PE count on \(j\)th SiPM within the time
interval \(\left[\underline{T}, \overline{T}\right]\) follows Poisson distribution
(Fig.~\ref{fig:explain_R}) of expectation
\begin{equation}
\lambda_{j, \left[\underline{T}, \overline{T}\right]}(\vec{r}, E) = \int_{\underline{T}}^{\overline{T}}
R_j \left(t;\vec{r}, E\right) \mathrm{d} t.
\label{eq:lambda}
\end{equation}
The ionization quenching and Cherenkov radiation~\cite{ADEY2019230} cause
the non-linearity between the number of emitted photons and the
kinetic energy of the charged particle.  Such \emph{physics non-linearity} is usually
modeled empirically and calibrated with monoenergetic sources, for example at Daya Bay~\cite{ADEY2019230}, RENO~\cite{PhysRevLett.112.061801} and Borexino~\cite{PhysRevLett.116.211801}.
In the scope of event reconstruction, \(E\) is measured in a scale proportional to the number of emitted photons,
also known as \emph{visible energy}, \(v(E)\).
Because \(v(E)\) describes photon generation, it is decoupled from photon propagation and detection in \(R_j(\cdot)\), resulting in separation of variables \(E\) and \(\vec{r}\),
\begin{equation}
 R_j(t;\vec{r}, E) = \underbrace{v(E)}_{\substack{\text{physics} \\ \text{non-linearity}}} \cdot \underbrace{R_j^0(t;\vec{r})}_{\substack{\text{geometric} \\\text{effect}}}.
 \label{eq:energy_linearity}
\end{equation}
\(R_j^0(t;\vec{r})\) encodes \emph{geometric effect},
the relative difference of the light curve over different \(\vec{r}\) at the \(j\)-th SiPM.

The good spherical symmetry of TAO CD makes the azimuth \(\phi\)
irrelevant in the relative position \((r,
\theta, \phi)\) between a vertex \(\vec{r}\) and position of the \(j\)th
SiPM \(\vec{r}_{\text{SiPM}, j}\)~(Fig.~\ref{fig:rel_p_def}).
After factoring out the quantum efficiency and time difference in the SiPM index \(j\),
for a vertex \(\vec{r}_i\), \(R_j^0(t; \vec{r}_i)\) merges into a single function \(R^0(t; r_i, \theta_{ji})\), where
\begin{equation}
r_i = \left|\vec{r}_i\right|,\quad
\cos \theta_{ji} = \left(\frac{\vec{r}_i \cdot \vec{r}_{\text{SiPM},j}}{
\left|\vec{r}_i\right| \left|\vec{r}_{\text{SiPM},j}\right|}\right).
\end{equation}

\begin{figure}
\centering
\begin{subfigure}{0.45\textwidth}
\centering
\begin{tikzpicture}[scale=0.7]
\begin{axis}[
    clip = false,
    axis lines = center,
    ticks = none,
    xlabel = \(t\),
    ylabel = \(R(t)\),
    xlabel style= below right,
    ylabel style= above left,
    legend style={font=\tiny, legend pos=outer north east,}
]
\addplot[
    name path=R,
    domain=0:5,
    samples=100
]{2.72^(-x/0.7) * (1 - 2.72^(-x / 5))};
\path [name path=axis] (axis cs:1, 0) -- (axis cs:2, 0);
\addplot
[
    pattern=north west lines
]
fill between[
    of=R and axis,
    soft clip={domain=1:2}
];
\node[below] at (100, 0) {\(T_1\)};
\node[below] at (200, 0) {\(T_2\)};
\node (lambda) at (350, 300) {\(\displaystyle \lambda = \int_{T_1}^{T_2} R(t) \mathrm{d}t\)};
\draw[thick, ->] (lambda) -- (150, 200);
\end{axis}
\end{tikzpicture}
\caption{}
\label{fig:explain_R}
\end{subfigure}
\begin{subfigure}{0.45\textwidth}
\centering
\begin{tikzpicture}[scale=1.1]
\draw (0, 0) node[below right] {\(O\)} circle (2);
\filldraw (0, 0) circle (0.03);
\draw[dotted, thick] (0,0) -- (1.01, 2.02);
\draw[dotted, thick] (0,0) -- node[below] {\(r\)} (-0.6, 0.6) node[below left] {\(\delta\left(\vec{r}, E\right)\)};
\draw[dotted, thick] (1.01, 2.02) -- (-0.6, 0.6);
\node [star, star point ratio=0.4, draw] at (-0.6,0.6) {};
\draw (0.1, 0.2) arc (26.6:170:0.15);
\draw (0.8, 1.6) arc (260:210:0.2);
\node at (-0.05, 0.5) {\(\theta\)};
\node at (0.55, 1.45) {\(\beta\)};
\draw[rotate=63.4, xshift=65] ellipse (0.1 and 0.25);
\node at (1.6, 1.7) {\(\vec{r}_{\text{SiPM}}\)};
\node at (0, 1.4) {\(l\)};

\end{tikzpicture}
\caption{}
\label{fig:rel_p_def}
\end{subfigure}

\caption{(\subref{fig:explain_R}) The physical meaning of response function \(R(t)\).  The
  PE count in \(\left[T_1, T_2\right]\) follows Poisson distribution,
  and the mean PE count is
  \(\lambda = \int_{T_1}^{T_2} R(t) \mathrm{d}t\).  (\subref{fig:rel_p_def}) The schematic diagram
  of relative positions \((r, \theta)\) of event vertex \(\vec{r}\) and
  SiPM \(\vec{r}_{\text{SiPM}}\) in CD.  \(\beta\) is the incident angle
  on SiPM.  \(l\) is the distance from vertex to the position of
  SiPM. The origin of spherical coordinate system \(O\) is put at the
  center of CD.  The detector is approximately symmetric
  rotationally about the \(O\)-\(\vec{r}_{\text{SiPM},j}\) axis, thus
  the relative azimuth \(\phi\) is ignored.}
\end{figure}

In TAO, the detector size is much smaller than the scattering or absorption lengths.
The variables of \(R(t; r, \theta)\) can be separated into time \(t\) and position \((r, \theta)\).

\subsection{Position part}
The response intensity \(\displaystyle{I(r, \theta) \coloneq \int R^0(t; r, \theta) \mathrm{d}t} \) is defined as the relative PE count on
a SiPM tile for given \((r, \theta)\). The accuracy of it dominates the energy non-uniformity
of reconstructed events. \(I(r, \theta)\) is approximately proportional to the solid angle \(\Omega\) of SiPM measured
from vertex and the exponential attenuation of distance \(l\) from vertex to the
position of SiPM,
\begin{equation}
\begin{aligned}
  I(r, \theta)\ &\propto\ \Omega\ \cdot \exp(-l/l_0)\\
  &\propto\
\frac{\cos \beta (r, \theta)}{r^2 + r_{\text{SiPM}}^2 - 2rr_{\text{SiPM}}\cos\theta}
\cdot \exp(-l/l_0),
\end{aligned}
\label{eq:position_response_classic}
\end{equation}
where \(\beta(r, \theta)\) is the incident angle on SiPM shown in
Fig.~\ref{fig:rel_p_def} and \(l_0\) is the attenuation length.
At TAO, more accuracy in \(I(\cdot)\) is needed. We follow
W.~Dou~et~al.~\cite{Dou2022ReconstructionOP} to
characterize the response intensity with Zernike polynomials \(Z_n(r, \theta)\)~\cite{Niu_2022} which are orthonormal on the unit disk
\begin{equation}
  I(r, \theta) = \Omega\ \cdot \exp \left[\sum_{n=0}^{N_z} a_n Z_n(r/r_{\max}, \theta)\right],
\label{eq:position_response_TAO}
\end{equation}
where \(r_{\max}=\SI{0.9}{m}\) and \(N_{z}\) is the maximum order.
The exponential in Eq.~\eqref{eq:position_response_TAO} maintains positiveness of the intensity
and encodes both the solid angle and the exponential attenuation components in Eq.~\eqref{eq:position_response_classic}.
\begin{equation}
    I'(r, \theta) = \left[\sum_{n=0}^{N_z} a_n Z_n(r/r_{\max}, \theta)\right]^2,
\label{eq:position_response}
\end{equation}
is a radically data-driven form entirely determined by the experimental data,
without physical consideration \emph{a priori}.

\begin{figure*}
\centering
\begin{subfigure}{0.32\textwidth}
\centering
\includegraphics[height=0.2\textheight]{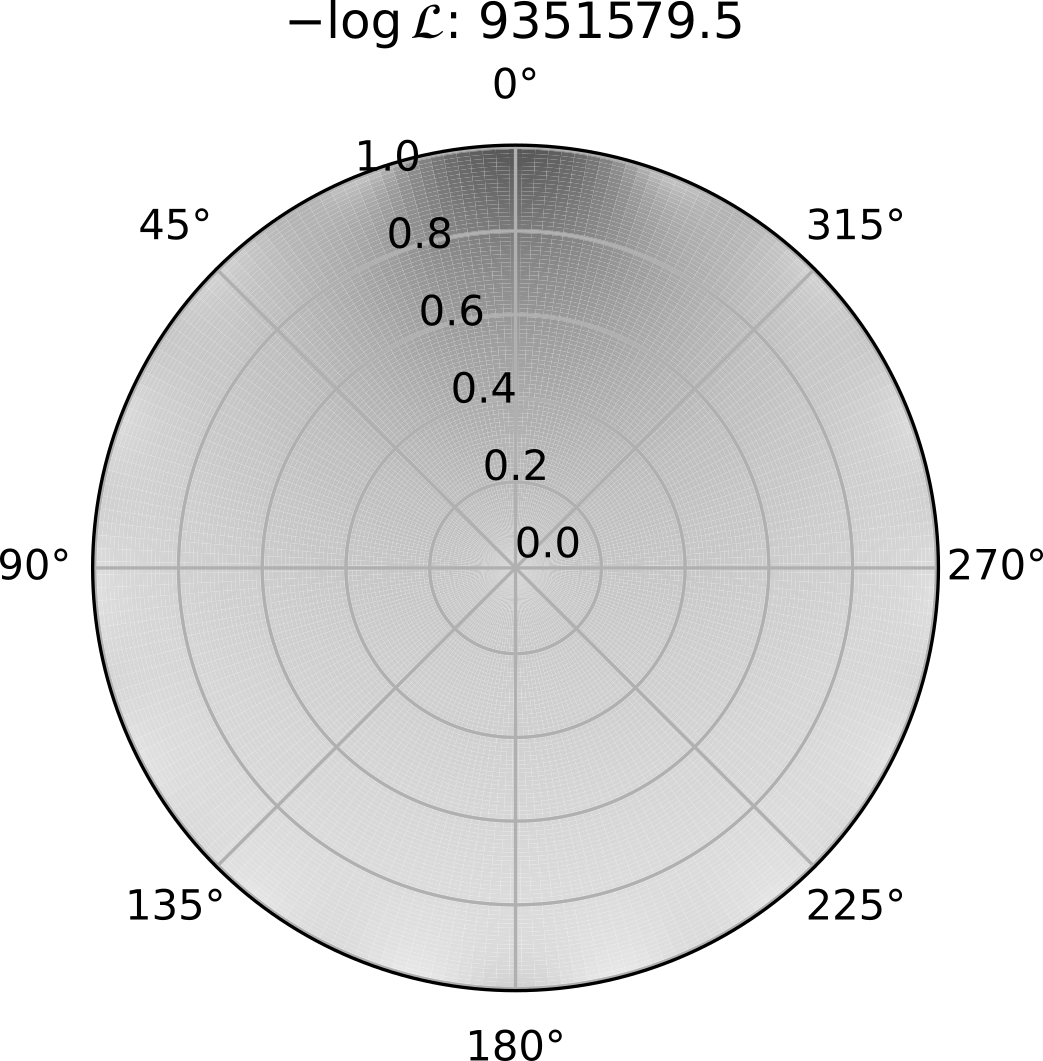}
\caption{Position response \(I'(r, \theta)\)}
\label{fig:tao_probe1_zernike}
\end{subfigure}
\hspace{-1.3em}
\begin{subfigure}{0.32\textwidth}
\centering
\includegraphics[height=0.2\textheight]{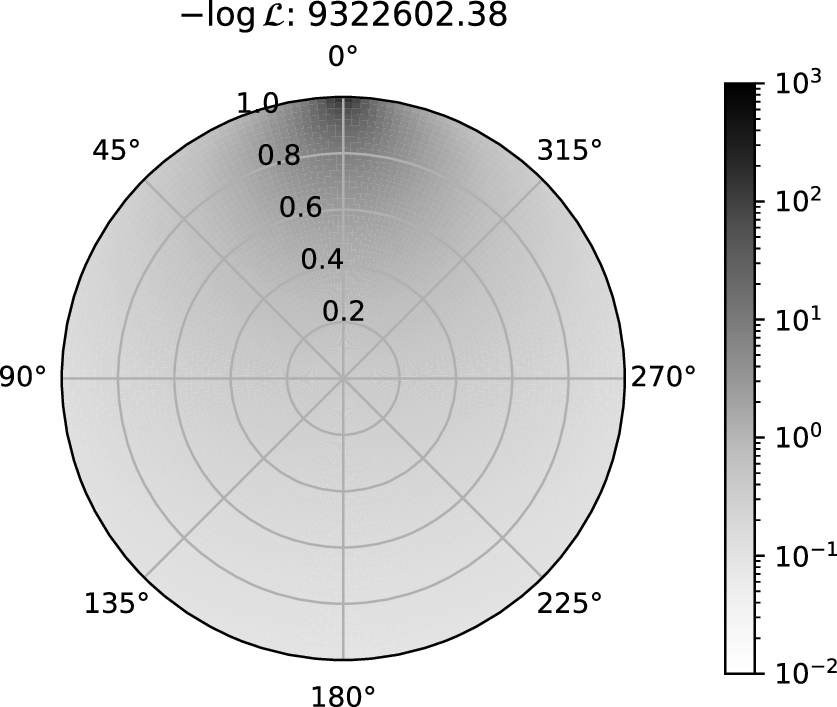}
\caption{Position response \(I(r, \theta)\)}
\label{fig:tao_probe1_invsquare}
\end{subfigure}
\begin{subfigure}{0.32\textwidth}
\centering
\includegraphics[width=\textwidth]{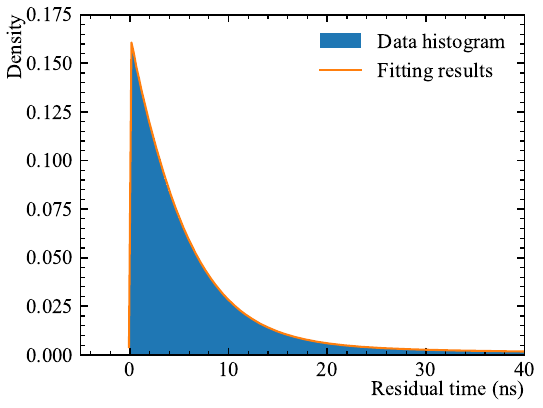}
    \caption{Time response \(P(t;r,\theta)\)}
\label{fig:tao_probe2}
\end{subfigure}
\caption{(\subref{fig:tao_probe1_zernike}) and (\subref{fig:tao_probe1_invsquare}) show the position response
    in \(I'(r, \theta)\) of square Zernike polynomials and \(I(r, \theta)\) of
    exponential geometrical construction.
    Two forms have the same order of Zernike polynomials.
    Owing to the consideration of solid angle in
    Eq.~\eqref{eq:position_response_TAO}, (\subref{fig:tao_probe1_invsquare})
    shows better description of rapidly changing intensity near the SiPM than
    (\subref{fig:tao_probe1_zernike}).
    The score (log-likelihood of the validation dataset) also indicates \(I(r,
    \theta)\) is better.
    (\subref{fig:tao_probe2}) shows time response \(P(t;r,\theta)\) fitted by
    Eq.~\eqref{eq:time_response} with 80-order Legendre polynomials. The bin width
    of histogram is \SI{0.2}{ns}.}
\end{figure*}

To decide
which form of the position response to use, we fit the same training
dataset with
Eqs.~\eqref{eq:position_response}\eqref{eq:position_response_TAO} and
use the same validation dataset introduced in Sec.~\ref{subsec:coefficient_fitting} to
evaluate them.
The log-likelihoods of Fig.~\ref{fig:tao_probe1_zernike} and Fig.~\ref{fig:tao_probe1_invsquare} indicate that the exponential of Zernike polynomials
is more suitable for the description of the position response.

\subsection{Time part}
\label{subsubsec:time_response}
\begin{figure}[htbp]
  \centering
  \includegraphics[width=0.4\textwidth]{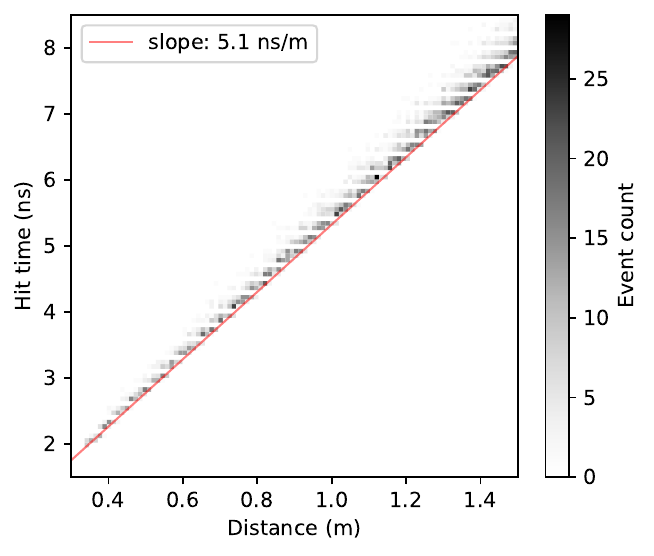}
  \caption{The histogram of hit times and distances. A linear equation (red line)
    is used to fit the lower edge of the histogram. The reciprocal of
    slope is the effective light speed.}
  \label{fig:dist_hittime_hist}
\end{figure}

We set the event times to 0 without loss of generality.
The separation of position and time variables implies that the shape
of the time response remains consistent
across all SiPM tiles and vertices in the CD.  To align the arrival times of
photons on different SiPM, we define the shift \(t_{\rm shift}\)
as the time of flight from vertex
\(\delta(\vec{r}, E)\) to the position of SiPM
\begin{equation}
t_{\rm shift}(r, \theta)  = \frac{n_{\text{LS}}l(r,\theta)}{c},
\label{eq:time_shift}
\end{equation}
where \(n_{\text{LS}}\) is the effective refractive index of liquid
scintillator (LS), \(l\) is the distance from vertex to the position of
SiPM and \(c\) is the speed of light in vacuum.
Scintillation photons often undergo changes in direction due to optical effects
such as absorption/re-emission, scattering, refraction, and reflection, making
their trajectory modeling challenging in detector response. A practical
approximation models the optical path as a straight line from the vertex to the
SiPM, as described in Eq.~\eqref{eq:time_shift}, a method validated in Z.~Li's
work~\cite{ziyuan2021}. The residual detector response can be calibrated using
polynomial functions.
To determine \(n_{\text{LS}}\),
we simulate \num[group-separator={,}]{10000} \SI{5}{MeV} electrons located at a
fixed point and get the peak position of hit time distribution for each SiPM.
Fig.~\ref{fig:dist_hittime_hist} shows a 2-D histogram of PE hit times
and distances \(l(r, \theta)\) on 4024 SiPM tiles.
The lower edge of the histogram represents the first-PE time and is linearly fitted to extract \(n_{\text{LS}}\).

The family of Legendre polynomial~\cite{ARFKEN2013715} is orthonormal on \([-1, 1]\).
We scale the residual time
\(t - t_{\text{shift}}\) to \([-1, 1]\) with
\begin{equation}
  t_{\text{scale}} = \frac{t - t_{\text{shift}} - \underline{T}}{\overline{T} -
    \underline{T}} \times 2 - 1,
\end{equation}
where \([\underline{T}, \overline{T}]\) is the residual time windows including
dominant part of time response,
\begin{equation}
P(t;r,\theta) = \exp\left[\sum_{m=0}^{N_p} a_m P_m(t_{\text{scale}})\right],
\label{eq:time_response}
\end{equation}
where \(P_m(\cdot)\) is the \(m\)-th order Legendre polynomial and
\(N_{p}\) is the maximum order of 80 in our test.
This value may vary for actual data but can be effectively evaluated using the
scoring funcion Eq.~\eqref{eq:calibration_function} to determine an optimal model.
The exponential is to ensure the time part to be nonnegative.

We simulate \num{100000} \SI{0.5}{MeV} (kinetic energy) electrons
distributed uniformly in CD, calculate the relative
positions for all PE hits in Fig.~\ref{fig:rel_p_def} and fit the coefficients \(a_m\) in Eq.~\eqref{eq:time_response}.
Fig.~\ref{fig:tao_probe2} shows the fitting results.
The residual time window \([\underline{T}, \overline{T}]\) is set to \([\SI{-0.5}{ns},
\SI{50}{ns}]\). The time response \(P(t)\) outside \([\underline{T},
\overline{T}]\) is considered as zero.
The maximum order of Legendre polynomials is determined by an independent
validation dataset.

The optical response function is
\begin{equation}
  R^0(t;r, \theta) =
  I(r, \theta) \cdot P(t;r,\theta).
  \label{eq:pmt_response_function}
\end{equation}

\subsection{Coefficients fitting and scoring}
\label{subsec:coefficient_fitting}

Electron is an ideal point source in LS because its energy deposition
occurs within a radius of a few millimeters~\cite{david2022}.
It deposits energy in CD and excites LS molecules.  The molecules
de-excite and emit scintillation photons, which transmit through the detector
and reach the SiPM to produce a PE in part. Those are simulated with
\textsc{Geant4}-based~\cite{allison_recent_2016} program. In the
simulation for coefficients fitting and
scoring, \num{100000} electrons with energy
\SI{0.5}{MeV} are distributed uniformly in the CD.
We fit parameters and select models in Eq.~\eqref{eq:pmt_response_function} by the likelihood
\begin{equation}
  \small
  \begin{aligned}
    \log \mathcal{L} =& \log\biggl\{\prod_{k} R^0 \left(t_{k}; r_{i_k},
             \theta_{j_ki_k}\right)\\
            &\cdot \prod_{i,j} \exp\Bigl[-\int R^0 (t; r_i,
             \theta_{ji}) \mathrm{d}t\Bigr] \biggr\}\\
           =& \underbrace{\sum_{k} \log R^0 \left(t_{k}; r_{i_k},
             \theta_{j_ki_k}\right)}_{\text{time part}} - \underbrace{\sum_{i,j}\int R^0 \left(t;
             r_{i}, \theta_{ji}\right) \mathrm{d}t}_{\text{PE part}},
  \end{aligned}
  \label{eq:calibration_function}
\end{equation}
where \(i\), \(j\) and \(k\) are indices of the event,
SiPM and PE. The ``PE part'' includes all the events and SiPMs, while the ``time part''
contains PE times \(t_k\) with their corresponding events \(i_k\) and SiPMs \(j_k\).

\section{Tweedie electronic time-charge likelihood} \label{sec:TweedieGLM}

Tan~\cite{10150/612894} formulates the PDF of single electron response~(SER) charge distribution in a Gaussian \(f_\mathrm{N}(Q; \mu, \sigma^2)\)
and the PE count \(N_{\text{PE}}\) in Poisson \(\pi(\lambda)\) where \(\lambda\) is the expectation.
The charge PDF of SiPM or PMT is:
\begin{equation}
  \begin{aligned}
    p(Q; \lambda, \mu, \sigma^2) = \sum_{N_{\rm PE}=0}^{\infty} &f_\mathrm{N}
    (Q; N_{\rm PE} \mu, N_{\rm PE} \sigma^2) \\
                                  &\cdot p_\pi(N_{\rm PE}; \lambda).
  \label{eq:simple_likelihood}
  \end{aligned}
\end{equation}
Although widely followed, it makes no physical sense for the Gaussian distribution
to allow a negative charge.
We follow Kalousis et al.~\cite{Kalousis_2020} to use a Gamma distribution \(\varGamma\left(k, \theta\right)\) to
model the SER charge distribution, where \(k\) and \(\theta\) are the shape and scale parameters.
Therefore, the distribution of total charge \(Q\)
\begin{equation}
  f_{\Tw}(Q; \lambda, k, \theta) = \sum_{N_{\rm PE}=0}^{\infty} f_{\varGamma}(Q; N_{\rm PE} k, \theta) p_\pi(N_{\rm PE}; \lambda),
  \label{eq:gamma_likelihood}
\end{equation}
follows compound Poisson-Gamma distribution. It is a special case of the Tweedie distribution~\cite{tweedie1984index}
where the Tweedie index parameter \(\xi\) satisfies \(1 < \xi <
2\)~\cite{poissongamma1996}.
Tweedie distribution includes the fluctuation of PE count,
thus the infinite \(N_{\text{PE}}\) summation
in Eq.~\eqref{eq:simple_likelihood} is shifted to standard routines~\cite{dunn_series_2005,dunn_evaluation_2008}.
The parameter relationship between Tweedie distribution
and its corresponding Poisson and Gamma distribution\cite{dunn_generalized_2018} is:
\begin{equation}
  \left\{
  \begin{aligned}
\lambda =& \frac{\mu^{2 - \xi}}{\phi(2 - \xi)}  \\
k =& \frac{2 - \xi}{\xi - 1}\\
\theta =& \phi \left(\xi - 1\right) \mu^{\xi - 1},\\
\end{aligned}\right.
\label{eq:pars_convert}
\end{equation}
where \(\mu\) and \(\phi\) are the mean value and dispersion parameters
of Tweedie distribution.

\subsection{Parameter calibration}
\label{subsubsec:coef_reg}
Tweedie distributions is a special case of exponential dispersion models (EDM)~\cite{jvarphirgensen1987exponential}.
Generalized linear model~(GLM)~\cite{mccullagh_generalized_1989,dunn_generalized_2018} is available for Eq.~(\ref{eq:gamma_likelihood}) to
establish the relationship between the expected PE count \(\lambda\)
and charge \(Q\).
Specifically, we use the following expression of GLM,
\begin{equation}
\begin{cases}
Q \sim \Tw(\mu, \phi, \xi)\\
\mu = b \lambda,
\end{cases}
\label{eq:TweedieGLM}
\end{equation}
with an identity link function \(g(\mu) = \mu\). The intercept of linear predictor is zero.
\(\lambda\) is predicted by the optical response from Eq.~(\ref{eq:position_response_TAO})
as the input to GLM. According to Eq.~\eqref{eq:pars_convert},
\begin{equation}
    \lambda k \theta = \mu \xrightarrow{\mu = b \lambda}
    b = k\theta = \E[Q | N_{\rm PE} = 1],
\end{equation}
the slope \(b\) is the expected charge of a single PE.

\begin{figure}[htbp]
  \centering
  \begin{subfigure}{0.4\textwidth}
  \includegraphics[width=\textwidth]{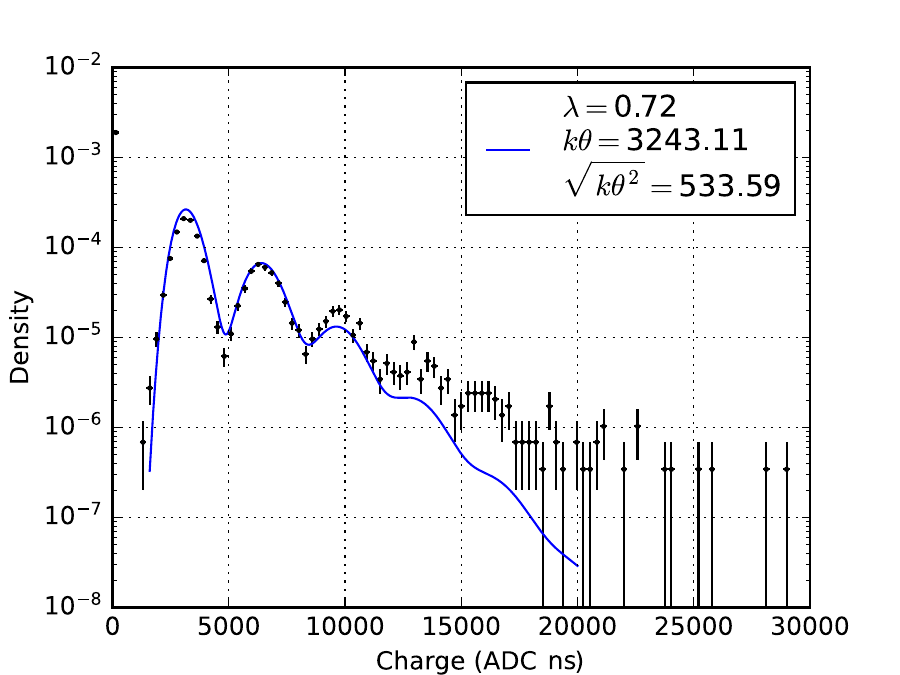}
  \caption{}
  \label{subfig:tweedie_charge_small_lambda}
  \end{subfigure}
  \begin{subfigure}{0.4\textwidth}
  \includegraphics[width=\textwidth]{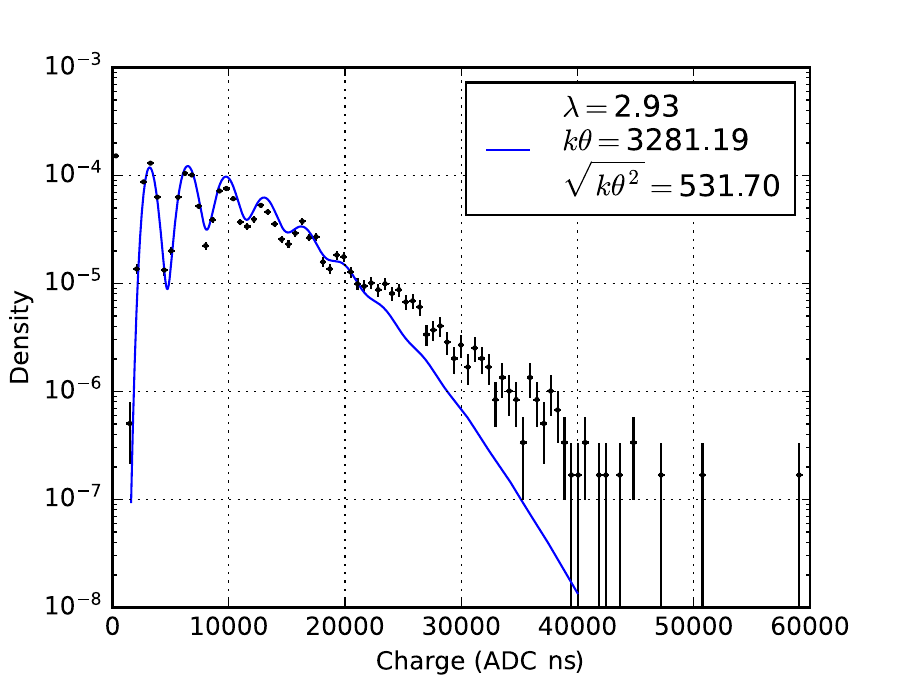}
  \caption{}
  \label{subfig:tweedie_charge_large_lambda}
  \end{subfigure}
  \begin{subfigure}{0.4\textwidth}
  \includegraphics[width=\textwidth]{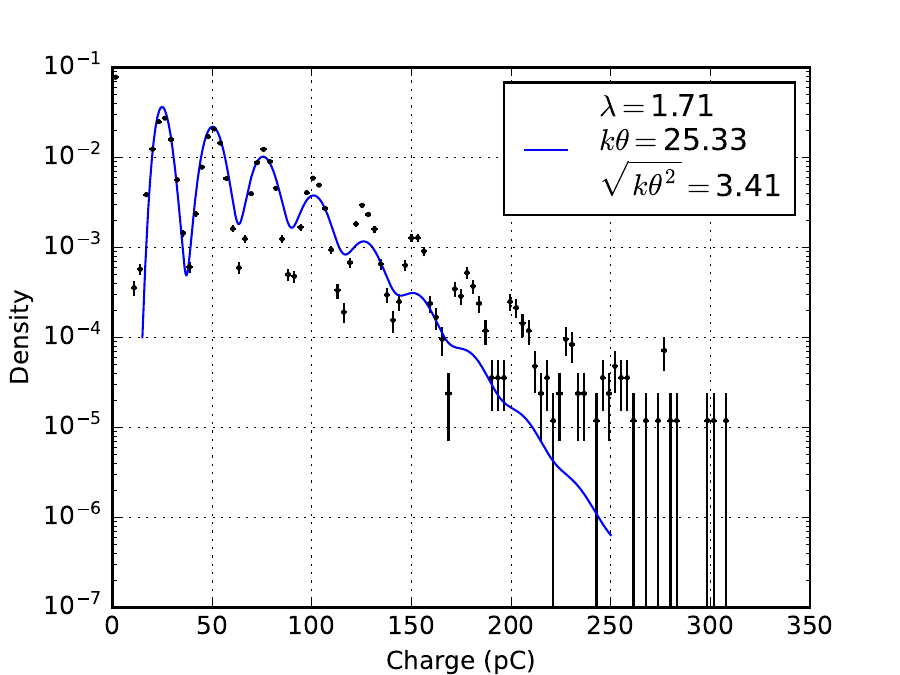}
  \caption{}
  \label{subfig:tweedie_charge_real}
  \end{subfigure}
  \caption{(\subref{subfig:tweedie_charge_small_lambda}) and
    (\subref{subfig:tweedie_charge_large_lambda}) are charge distributions
    of a SiPM tile for
    \num[group-separator={,}]{10000} \SI{1}{MeV} and \SI{3}{MeV} electrons fixed at
    center of CD (Monte Carlo
    simulation). (\subref{subfig:tweedie_charge_real}) is the charge distribution
    from one channel output of SiPM tile (experimental data). Solid lines
    are the regression results of Tweedie GLM, where the parameters of
    Poisson (\(\lambda\)) and Gamma (\(k\) and \(\theta\)) parts are listed.
    The units of charge have a conversion factor between \si{ADC\ ns}
    and \si{pC} that do not challenge the validity of the Tweedie model.}
  \label{fig:charge_distribution}
\end{figure}

For simplicity,
we ignore the variations of the SiPM-tile Tweedie parameters in
the Monte Carlo.  In the future we shall calibrate the real detector channel-by-channel.
Figs.~\ref{subfig:tweedie_charge_small_lambda}
and \ref{subfig:tweedie_charge_large_lambda} show the charge distribution of a
selected SiPM for \num[group-separator={,}]{10000} \SI{1}{MeV} and \SI{3}{MeV}
electrons located at the center of CD, where \(\lambda\) is kept constant.
These charges are generated by \emph{electronic simulation}
considering dark noise, afterpulse and internal
crosstalk~\cite{ACERBI201916}.

Our electronic simulation includes internal crosstalk,
where every PE might induce another PE in the SiPM.
It breaks the Poisson assumption in Eq.~(\ref{eq:gamma_likelihood}) and necessitates
a generalized Poisson~\cite{doi:10.1080/03610929208830766,VINOGRADOV2012247} suggested by Vinogradov~\cite{vinogradov_analytical_2012},
with a probability mass function~(PMF) of
\begin{equation}
f_{\text{GP}} (x;\theta, \eta) = \frac{\theta (\theta + \eta x)^{x - 1}
  \mathrm{e}^{-\theta - \eta x}}{x!}.
\label{eq:GP}
\end{equation}
It is verified to work in V.~Chmill~\cite{CHMILL201770} and
Jack~Rolph~\cite{ROLPH2023168544}'s studies.

Although when the crosstalk rate \(\eta\rightarrow0\) Eq.~\eqref{eq:GP} degenerates
back to a Poisson, the extended compound distribution is generally not in the Tweedie family any more.
Fortunately, when \(\lambda\) is not much larger than \(1\) and
the probability of crosstalk is as low as \SI{\sim 15}{\percent},
the effect is not serious.
The Tweedie model of Eq.~(\ref{eq:gamma_likelihood}) is validated against a laboratory test of a SiPM sample in
Fig.~\ref{subfig:tweedie_charge_real}.  The data and model exhibits difference
at the higher charge tails of the distribution, similar to the Monte Carlo
in Fig.~\ref{subfig:tweedie_charge_large_lambda}.
Momentarily, we regard the convenience of Tweedie GLM to surpass the imperfectness of the Poisson assumption,
as will be supported by the reconstruction results in Sec.~\ref{sec:dataset}.
We shall develop regression with the compound of generalized Poisson and Gamma distribution
in our future publications.


\subsection{Charge-only reconstruction likelihood}
\label{subsec:charge_rec_lik}

For some small detectors, time readout is usually not important.
The expected PE count in the electronic time window \(\left[\underline{T},
  \overline{T}\right]\) is
\begin{equation}
  \begin{aligned}
    \lambda_{j, \left[\underline{T}, \overline{T}\right]}(\vec{r}, E)
    = & v(E) \int_{\underline{T}}^{\overline{T}} R_j^0(t;\vec{r}) \mathrm{d}t   \\
    = & v(E) \lambda_{j,\left[\underline{T}, \overline{T}\right]}^{0}(\vec{r}).
  \end{aligned}
\end{equation}
A charge-only reconstruction likelihood is a direct consequence of
the Tweedie distribution in Eq.~(\ref{eq:gamma_likelihood}) and the optical model in Eq.~(\ref{eq:energy_linearity}),
\begin{equation}
  \small
  L(\vec{r}, v_{E}; \{Q_j\}) = \prod_{j=1}^{N_{\text{SiPM}}} f_{\Tw}(Q_j;bv_{E}\lambda^{0}_{j,\left[\underline{T}, \overline{T}\right]}(\vec{r}), \phi, \xi),
  \label{eq:charge_lik}
\end{equation}
where \(b\), \(\phi\) and \(\xi\) are calibrated before event
reconstruction. \(v_{E}\) is an alternative notation for \(v(E)\) just to remind us
that visible energy is the target of the likelihood-based estimation. It is \(v(E)\)
instead of the kinetic energy \(E\) that needs to be treated as a parameter.

The dark counts of SiPM follow a homogeneous Poisson process with a constant
intensity \(R_{\text{D}}\) in our Monte Carlo simulations. The simulation
utilizes a dark rate of \SI{20}{Hz/mm^2}, which is derived from SiPM mass
testing and assumed uniform for all SiPM tiles. Each SiPM tile consists of 32
pixels measuring \(12 \times 6\)~\si{mm^2}, resulting in an aggregate dark rate of
\(R_{\text{D}}=20 \times 32 \times 12 \times 6 = 46048~\text{Hz}\). However, this value is not
definitive for actual detectors, as the dark rate is affected by the SiPM's
operating temperature and bias voltage.
By the superposition property of Poisson
process, it is incorporated into the optical model at Eq.~(\ref{eq:charge_lik})
by replacing \(R_j(\cdot)\) with \(R_j(\cdot) + R_{\text{D}}\)
\begin{equation}
  \begin{aligned}
    \lambda_{j,\left[\underline{T}, \overline{T}\right]}(\vec{r}, E)
    = & \int_{\underline{T}}^{\overline{T}}\bigl[ R_j(t;\vec{r}, E) +
    R_{\text{D}}\bigr]\mathrm{d}t                                                                                                  \\
    = & v(E)\lambda_{j,\left[\underline{T}, \overline{T}\right]}^{0}(\vec{r}) + R_{\text{D}} \cdot (\overline{T} - \underline{T}).
  \end{aligned}
  \label{eq:dn}
\end{equation}

\subsection{Time-charge reconstruction likelihood}
\label{subsec:TQ_distribution}

The time-charge reconstruction is challenging to get correct because of
the inter-dependence between the two variables.  The charge \(Q\) affects
first hit time \(T\) indirectly via the PE counts, not to be confused with the
time-walk~\cite{Kurtti2009PulseWT} which is a time error caused
by varying amplitude of pulses and a constant threshold.
Conversely, an observed \(T\) implies
the integrated charge is only contributed by the time window of \(\left[T, \overline{T}\right]\).
It invalidates all the prevents efforts trying to decouple the time-charge reconstruction likelihood
into time and charge parts.  Instead, we should start from first principles to derive the
joint distribution of \(T\) and \(Q\).

For clarity in the following derivation, we write \(R(t)\) to mean \(R_j(t; \vec{r}, E)\).
Consider the following two events. \emph{Event A}: There is no PE or charge in \(\left[\underline{T}, T\right]\).
\emph{Event B}: There is no charge in \(\left[\underline{T}, T + \Delta T\right]\) (\(\Delta T > 0\)).
Obviously, \(B \subset A\). Fig.~\ref{fig:eventAB} shows the probabilities of both events.
\begin{figure*}
  \centering
  \begin{tikzpicture}[
      declare function={gamma(\z)=
          (2.506628274631*sqrt(1/\z) + 0.20888568*(1/\z)^(1.5) + 0.00870357*(1/\z)^(2.5) - (174.2106599*(1/\z)^(3.5))/25920 - (715.6423511*(1/\z)^(4.5))/1244160)*exp((-ln(1/\z)-1)*\z);},
      declare function={gammapdf(\x,\k,\theta) = \x^(\k-1)*exp(-\x/\theta) / (\theta^\k*gamma(\k));}
    ]
    \begin{axis}[
        axis lines=none,
        enlargelimits=upper,
        height=5cm,
        xshift=3cm
      ]
      \addplot [smooth, domain=0:10, dotted] {gammapdf(x,2,1)};
    \end{axis}

    \draw [ultra thick] (0, 0) -- (10, 0);
    \draw [thick, ->] (-0.5, 0) -- (10.5, 0);
    \draw [blue,thick] (0, 0) -- (0, 0.3);
    \draw [blue,thick] (10, 0) -- (10, 0.3);
    \draw [blue,thick] (4, 0) -- (4, 0.3);

    \draw [red,thick] (0, 0) -- (0, -0.3);
    \draw [red,thick] (10, 0) -- (10, -0.3);
    \draw [red,thick] (4.5, 0) -- (4.5, -0.3);

    \draw [decorate,decoration={brace, amplitude=5},thick,blue] (0, 0.4)
    -- (4, 0.4) ;
    \draw [decorate,decoration={brace, amplitude=5},thick,blue] (4, 0.4)
    -- (10, 0.4) ;
    \draw [decorate,decoration={brace, amplitude=5, mirror},thick,red] (0, -0.4)
    -- (4.5, -0.4) ;
    \draw [decorate,decoration={brace, amplitude=5, mirror},thick,red] (4.5, -0.4)
    -- (10, -0.4) ;

    \node at (10.3, -0.3) {\(t\)};
    \node at (-0, 0.8) {\(\underline{T}\)};
    \node at (10, 0.8) {\(\overline{T}\)};
    \node at (4, 0.8) {\(T\)};
    \node at (4.5, -0.8) {\(T+\Delta T\)};
    \node at (4.25, 2.2) {\(R(t)\)};

    \node [blue] at (-1.0, 0.4) {Event A:};
    \node [red] at (-1.0, -0.4) {Event B:};

    \node [blue] at (2, 1.2) {\(\exp\left[-\int_{\underline{T}}^{T} R(t)
        \mathrm{d}t\right]\)};
    \node [blue] at (7, 1.2) {\(\left.f_{\Tw} \left(Q ;
    \lambda\right)\right|_{\lambda = \int_{T}^{\overline{T}} R(t) \mathrm{d}t}\)};
    \node [red] at (2, -1.2) {\(\exp\left[-\int_{\underline{T}}^{T+\Delta T} R(t)
        \mathrm{d}t\right]\)};
    \node [red] at (7.5, -1.2) {\(\left.f_{\Tw} \left(Q ;
    \lambda\right)\right|_{\lambda =\int_{T+\Delta T}^{\overline{T}} R(t) \mathrm{d}t}\)};
  \end{tikzpicture}
  \caption{Diagram of response function \(R(t)\) (dotted line), event A (blue) and
    event B (red) along the time axis.  Event A contains Event B due to the
    one-way dimension of object time \(t\) and \(\Delta T > 0\).  The
    probabilities of each sub events are listed.  }
  \label{fig:eventAB}
\end{figure*}
The set difference \(A \setminus B\) has a physical meaning that
there is no charge in \(\left[\underline{T}, T\right]\), and there is a PE
in \(\left[T, T + \Delta T\right]\), and
the \(Q\) is generated by \(\int_{T}^{\overline{T}} R(t) \mathrm{d}t\).
The difference of their probabilities is
\begin{equation}
  {\small
    \begin{aligned}
        & f_{\text{TQ}}[T, Q ; R(t)] \Delta T                                                \\
      = & \overbrace{\exp\left[-\int_{\underline{T}}^{T} R(t) \mathrm{d}t\right]
      \left.f_{\Tw} \left(Q ; \lambda\right)\right|_{\lambda =
      \int_{T}^{\overline{T}} R(t) \mathrm{d}t}}^{\text{Event A}}                            \\
        & - \underbrace{\exp\left[-\int_{\underline{T}}^{T+\Delta T} R(t) \mathrm{d}t\right]
      \left.f_{\Tw} \left(Q
      ;\lambda\right)\right\vert_{\lambda = \int_{T+\Delta T}^{\overline{T}} R(t)
      \mathrm{d}t}}_{\text{Event B}}. \label{eq:TQ_difference}
    \end{aligned}
  }
\end{equation}
When \(\Delta T \rightarrow 0\),
\begin{equation}
  {\small
    \begin{aligned}
        & f_{\text{TQ}}[T, Q; R(t)]                                                      \\
      = & -\frac{\partial}{\partial T}\left\{ \exp\left[-\int_{\underline{T}}^{T} R(t)
        \mathrm{d}t\right] \left. f_{\Tw} \left(Q ;
      \lambda\right)\right|_{\lambda = \int_{T}^{\overline{T}} R(t) \mathrm{d}t}\right\} \\
      = & \exp\Bigl[-\int_{\underline{T}}^{T} R(t)
      \mathrm{d}t\Bigr] R(T)                                                             \\
        & \cdot \left.\left(1 +
      \frac{\partial}{\partial \lambda}\right) f_{\Tw} \left(Q ;
      \lambda \right)\right|_{\lambda = \int_{T}^{\overline{T}} R(t) \mathrm{d}t}.
    \end{aligned}
  }
  \label{eq:TQ}
\end{equation}
Eq.~\eqref{eq:TQ} is the joint distribution of charge \(Q\) and first
hit time \(T\), whose normalization is verified in Appendix~\ref{appendix_A}.

When \(R(t)\equiv\rho\) is a constant, the time terms of
Eq.~\eqref{eq:TQ} resembles a random-start waiting time of a
paralyzable deadtime~\cite{doi:10.1093/jicru} with length
\(T-\underline{T}\),
\begin{equation}
  g(t) = \rho \exp[-\rho (T-\underline{T})].
\end{equation}
At TAO, any PE is only registered in a trigger-initiated data-taking
window. That is different from a nuclear counting circuit where
signals are continuously recorded.  Despite this, they do share the same
logic that a signal cannot be registered if there is another one in
the preceding deadtime interval.  Consequently, TAO electronics
appears to have a varied deadtime \(T-\underline{T}\) according to the
location of the first PE in the time window.

Expanding \(R(t)\) back to \(R_j(t; \vec{r}, E)\), the reconstruction likelihood is
\begin{equation}
  {\small
  \begin{aligned}
    &L\left( \vec{r}, v_{E}, t_0; \{(T_j,Q_j)\} \right)\\
    =& \prod_{\substack{Q_j > 0        \\\text{hit}}} f_{\text{TQ}}
    \left[T_j,Q_j; v_{E} R^{0}_j(t-t_0; \vec{r})\right] \\
     &\times \prod_{\substack{Q_j = 0     \\ \text{nonhit}}} p_{\pi}\left(0; v_{E} \lambda^{0}_{j,\left[\underline{T} - t_0, \overline{T} - t_0\right]}\left(\vec{r}\right) \right) \\
    = & \prod_{Q_j>0} \Biggl\{
    \exp\left[-v_{E}\lambda^{0}_{j,\left[\underline{T} - t_0, T_j -
    t_0\right]}\left(\vec{r}\right)\right]\\
      &\qquad\cdot v_{E} R^{0}_j\left(T_j - t_0;\vec{r}\right)\\
      &\qquad\cdot \left.\left(1 + \frac{\partial}{\partial
      \lambda}\right) f_{\Tw} \left(Q_j ;
    \lambda\right)\right|_{\lambda = v_{E}\lambda^{0}_{j,\left[T_j - t_0, \overline{T} - t_0 \right]}\left(\vec{r}\right)
    } \Biggr\}                          \\
      & \times \prod_{Q_j=0} \exp\left[
    - v_{E}\lambda^{0}_{j,\left[\underline{T} - t_0, \overline{T} - t_0\right]}\left(\vec{r}\right) \right],
  \end{aligned}
  }
  \label{eq:fht_charge_lik}
\end{equation}
where \(t_0\) is the event time and \(j\) is the index of SiPM.
Inclusion of dark hits is straightforward by substituting \(R_j(\cdot)\) with \(R_j(\cdot) + R_{\text{D}}\)
as Eq.~\eqref{eq:dn}.


\section{Numerical experiment}
\label{sec:dataset}

Calibration runs with radioisotopes~\cite{xu_calibration_2022} will
be the benchmark for event reconstruction.  Before such data are available,
we deploy Monte Carlo simulation to fit the coefficients of response function
and evaluate the reconstruction.

Generally, the simulation is carried out in two stages. \emph{Detector}
and \emph{electronic simulation} cover the processes before and after
a photon hits a SiPM.  The initial velocities of the electrons are isotropic
in \emph{detector simulation}. PEs, the information carriers for event
reconstruction,
are smeared in both number and times in \emph{electronic simulation}.
Table~\ref{table:dataset} summarizes the simulated datasets used to calibrate
the detector response and evaluate the reconstruction performance.
The \emph{detector simulation} of \num{100,000}
\SI{0.5}{MeV} electrons uniformly distributed within the CD are
employed to calibrate the optical model, as detailed in
Sec.~\ref{sec:detector_response}.  The \emph{detector} and
\emph{electronic simulation} of \num{10000} \SI{1}{MeV} and \SI{3}{MeV}
electrons located at the center of CD are for the parameter calibration of Tweedie
distribution in the electronic model, as discussed in Sec.~\ref{subsubsec:coef_reg}.
Finally, for the evaluation of reconstruction performance, electrons with fixed
energies and vertices along the \(x\)-axis are simulated, as will be discussed
in the following sections.
Owing to the high photon-coverage of nearly 94\% and uniform arrangement of SiPM
tiles, the optical response remains consistent within the fiducial volume and
the deviations from spherical symmetry is negligible.
The reconstruction results for electrons along the z-axis are consistent with
those obtained along the x-axis within a radius of \SI{700}{mm},
indicating that the reconstruction performance of x-axis events represents this
volume well.

\begin{table*}[htbp]
    \centering
    \caption{\ce{e-} datasets used for calibration of response and
    evaluation of reconstruction methodology}
    \label{table:dataset}
    \begin{tabular}{cccc}
    \toprule
    Usage & Section & Simulation stage
    & Configuration \\
    \midrule
    Optical model & \ref{sec:detector_response} & detector
    & \SI{0.5}{MeV} \ce{e-} uniformly in CD \\
    Electronic model & \ref{sec:TweedieGLM} & detector + electronic &
    \SI{1}{MeV} and \SI{3}{MeV} \ce{e-} at the detector center \\
    Evaluation & \ref{sec:dataset} & detector + electronic &
    \(0.5 \sim 7.5 \si{MeV}\) \ce{e-} along the x-axis \\
    \bottomrule
    \end{tabular}
\end{table*}

We access two variants of reconstruction likelihood functions in
Eqs.~\eqref{eq:charge_lik} and \eqref{eq:fht_charge_lik}. Both of them consider
the dark rate \(R_{\text{D}}\).

\subsection{With charge \(Q\)}

\begin{figure*}[htbp]
  \centering
  \begin{subfigure}{0.35\textwidth}
    \centering
    \includegraphics[width=\textwidth]{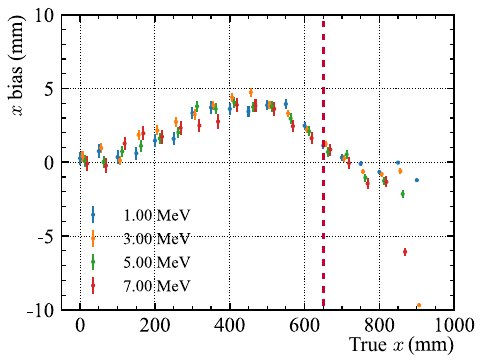}
    \caption{}
    \label{subfig:x_bias_Q}
  \end{subfigure}
  \begin{subfigure}{0.35\textwidth}
    \centering
    \includegraphics[width=\textwidth]{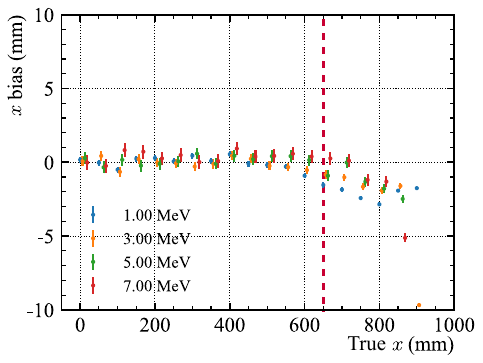}
    \caption{}
    \label{subfig:x_bias_Q_fht}
  \end{subfigure}\\
  \begin{subfigure}{0.35\textwidth}
    \centering
    \includegraphics[width=\textwidth]{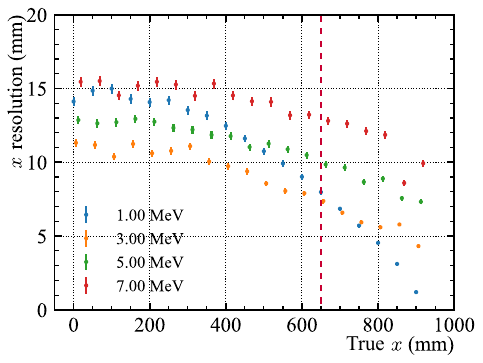}
    \caption{}
    \label{subfig:x_res_Q}
  \end{subfigure}
  \begin{subfigure}{0.35\textwidth}
    \centering
    \includegraphics[width=\textwidth]{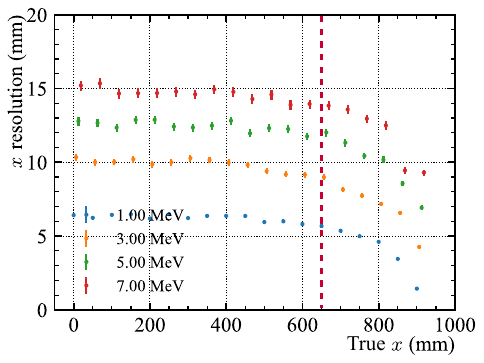}
    \caption{}
    \label{subfig:x_res_Q_fht}
  \end{subfigure}\\
  \begin{subfigure}{0.35\textwidth}
    \centering
    \includegraphics[width=\textwidth]{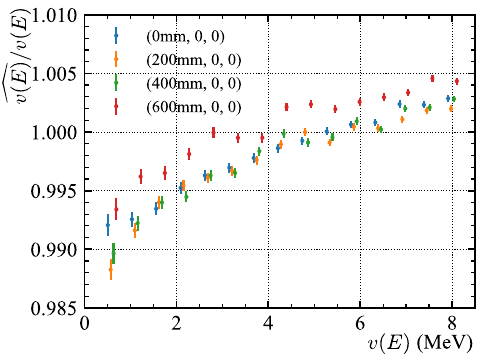}
    \caption{}
    \label{subfig:E_ratio_Q}
  \end{subfigure}
  \begin{subfigure}{0.35\textwidth}
    \centering
    \includegraphics[width=\textwidth]{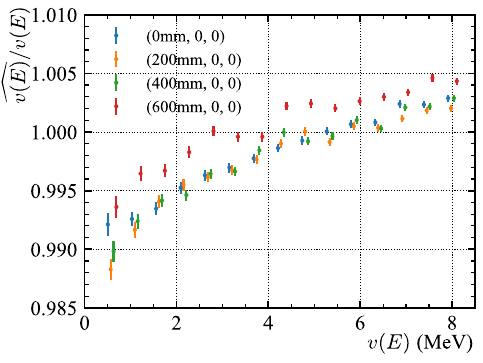}
    \caption{}
    \label{subfig:E_ratio_Q_fht}
  \end{subfigure}\\
  \begin{subfigure}{0.35\textwidth}
    \centering
    \includegraphics[width=\textwidth]{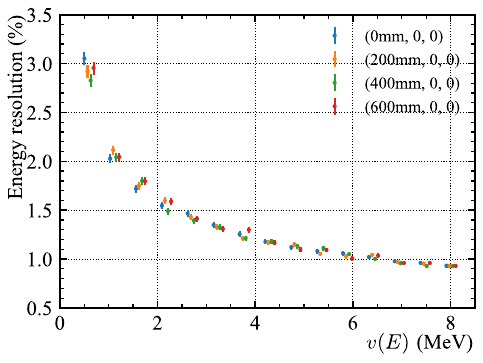}
    \caption{}
    \label{subfig:E_res_Q}
  \end{subfigure}
  \begin{subfigure}{0.35\textwidth}
    \centering
    \includegraphics[width=\textwidth]{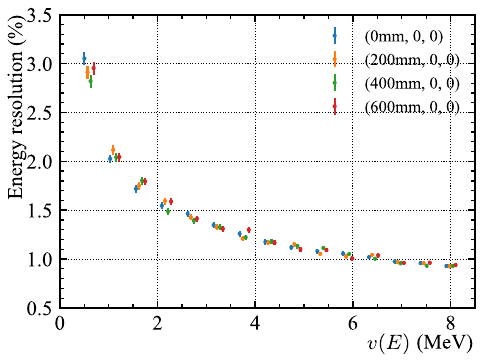}
    \caption{}
    \label{subfig:E_res_Q_fht}
  \end{subfigure}
  \caption{The reconstruction results of vertex position (\(x\) coordinates) and
    energy using charge (first column) and first hit time(second column).
    (\subref{subfig:x_bias_Q}), (\subref{subfig:x_bias_Q_fht}),
    (\subref{subfig:x_res_Q}) and (\subref{subfig:x_res_Q_fht})
    show the reconstruction bias
    and resolution (standard deviation) of \(x\) coordinates, with vertical dashed lines
    marking the boundary of fiducial volume;
    (\subref{subfig:E_ratio_Q}), (\subref{subfig:E_ratio_Q_fht}),
    (\subref{subfig:E_res_Q}), (\subref{subfig:E_res_Q_fht})
    show the reconstruction bias and resolution (relative standard
    deviation) of visible energy. The data points are shifted horizontally for visibility.}
  \label{fig:x_E_rec}
\end{figure*}

The first column of Fig.~\ref{fig:x_E_rec} shows the reconstruction by
charge using Eq.~\eqref{eq:charge_lik}.
Fig.~\ref{subfig:x_bias_Q} gives the bias of reconstructed vertices
along the \(x\)-axis.
The maximum bias in the fiducial volume (FV) is about \SI{5.0}{mm},
which occurs around radius of \SI{400}{mm}.
Vertices near the boundary of CD won't be mis-reconstructed into the FV.
The vertex bias is caused by the approximation of
intensity function Eq.~\eqref{eq:position_response_TAO}.
For vertex resolution shown in Fig.~\ref{subfig:x_res_Q},
we find it decrease with energy at \SI{<3}{MeV} but increase at \SI{>3}{MeV}.
At low energy, an electron deposits its energy within several millimeters.
The vertex resolution is determined by sheer quantity of photons.
At high energy, an electron travels centimeters long that is comparable to the vertex resolution.
The resolution gets worse with longer tracks.

We define the visible energy \(v(E)\) of an event as a linear scale
from the expected number of PEs \(\langle N_{\text{PE}}(E)\rangle = \langle \sum_{j=1}^{N_\text{SiPM}} p_j N_{\text{PE}, j}(E) \rangle\)
if it were at the detector center, with the conversion factor that
makes an electron with \SI{0.5}{MeV} kinetic energy the same visible energy
of \SI{0.5}{MeV},
\begin{equation}
  v(E) = \frac{\langle \sum_{j=1}^{N_\text{SiPM}} p_j N_{\text{PE}, j}(E) \rangle}{\langle
    \sum_{j=1}^{N_\text{SiPM}} p_j N_{\text{PE}, j}(\SI{0.5}{\MeV}) \rangle}\SI{0.5}{MeV},
  \label{eq:vE_defination}
\end{equation}
while \(p_j\) indicates the differences of photon detection efficiency~(PDE)
among SiPM tiles. In the simulation these PDEs are the same. We ignore the
\(p_j\) in \eqref{eq:vE_defination} and use the simplied form,
\begin{equation}
  \begin{aligned}
    v(E) =& \frac{\langle \sum_{j=1}^{N_\text{SiPM}} N_{\text{PE}, j}(E) \rangle}{\langle
    \sum_{j=1}^{N_\text{SiPM}} N_{\text{PE}, j}(\SI{0.5}{\MeV})
  \rangle}\SI{0.5}{MeV}\\
      =& \frac{\langle N^0_{\text{PE}}(E) \rangle}{\langle
    N^0_{\text{PE}}(\SI{0.5}{\MeV}) \rangle}\SI{0.5}{MeV}.
    \label{eq:vE_defination_0}
  \end{aligned}
\end{equation}

\begin{figure}[htbp]
  \centering
  \includegraphics[width=0.45\textwidth]{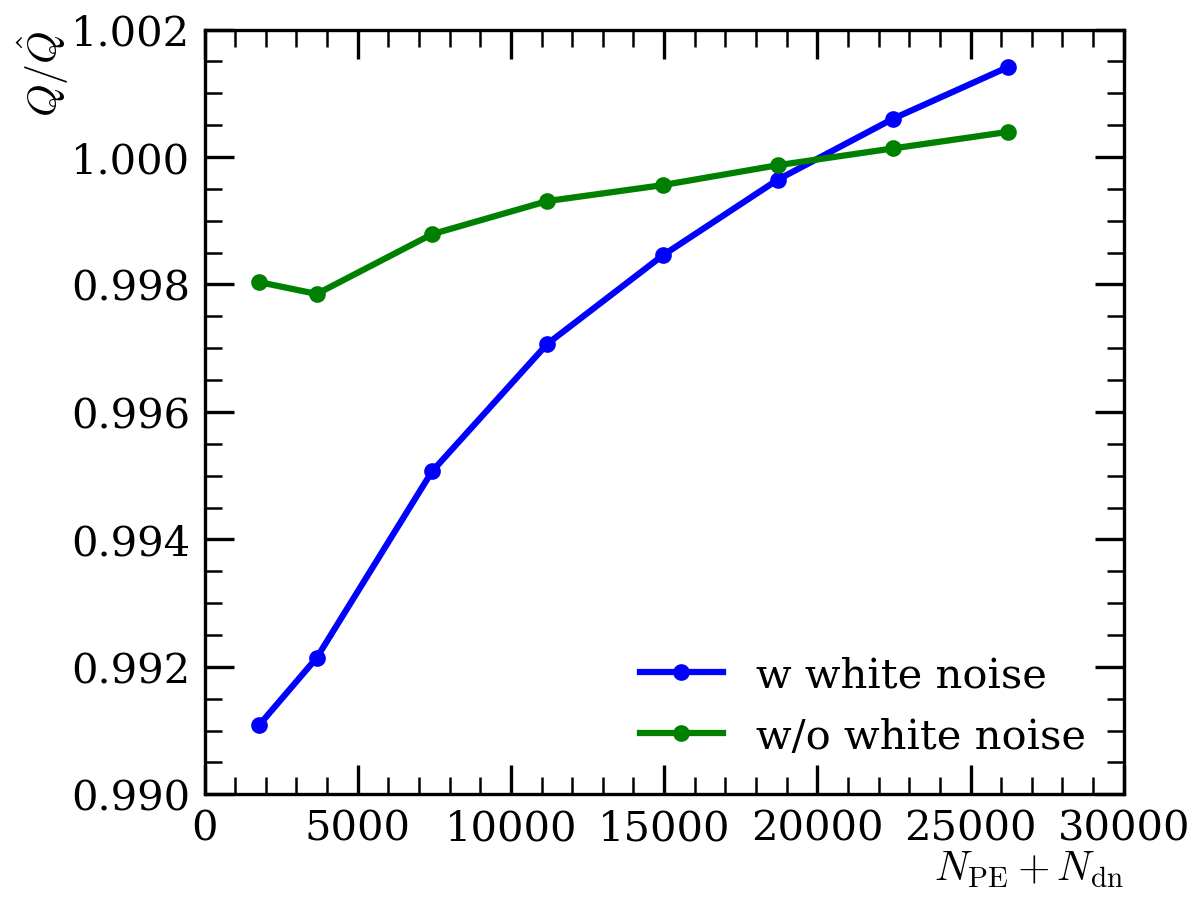}
  \caption{The ratio \(Q/\hat{Q}\) indicates the electronics non-linearity.
    The test data are electrons located at center of CD.}
  \label{fig:QADC_npe_rr}
\end{figure}

Fig.~\ref{subfig:E_ratio_Q} shows the ratio
of reconstructed energy \(\widehat{v(E)}\) versus visible energy
\(v(E)\), which is caused by the deviation from
linearity between PE count and output charge on SiPMs, known as
\emph{electronics non-linearity}.
To assess whether the Tweedie GLM introduces any additional non-linearity,
a linear fit was employed to predict charge \(Q\) from the PE count
\(N_{\text{PE}}\). \(N_{\text{PE}}\) contains physical PEs that is proportional
to visible energy \(v(E)\) and dark counts \(N_{\text{dn}}\).
This procedure simplifies the computational load during
checks, thereby avoiding the need for multiple reconstructions.
The ratio \(Q/\hat{Q}\) in Fig.~\ref{fig:QADC_npe_rr} matches
with Fig.~\ref{subfig:E_ratio_Q} in both shape and magnitude,
indicating the absence of extra non-linearity from Tweedie GLM.
Contribution from white noise dominates the electronics non-linearity.

Fig.~\ref{subfig:E_res_Q} shows the resolution of energy
reconstruction to be slightly over \SI{2}{\percent} at \SI{1}{MeV}.
We shall discuss it more around Eq.~\eqref{eq:Eres_abc} in the next section.

\subsection{With charge \(Q\) and accurate first hit time \(T\)}
\label{subsec:res_TQ}
In meter-scaled liquid scintillator detectors, the vertex resolution
is dominated by charges because the spread of scintillation time
profile is comparable to the photon time of flight~\cite{GALBIATI2006700}.
But time is useful for reducing reconstruction bias and pulse-shape discrimination in our next study.
In this section, the
first hit times are extracted from the detector simulation without
imposing electronic smears to evaluate its best possible contribution.

The right column of Fig.~\ref{fig:x_E_rec} shows the reconstruction
using charge and first hit time. The bias of vertex reconstruction
shown in Fig.~\ref{subfig:x_bias_Q_fht} increases with radius. In the
FV \(r < \SI{650}{mm}\), the maximum bias is \SI{2.0}{mm},
substantially less than that using only charge in
Fig.~\ref{subfig:x_bias_Q}.  The vertex resolution in
Fig.~\ref{subfig:x_res_Q_fht} is better than that using charge only
especially for low energy (\SI{\sim 1}{MeV}).
In the FV, it shows a new flat trend with time, because
the accuracy of times only degrades slightly due to dispersion when a
source moves away from SiPMs, much less sensitive than that of charges.
Our results show that time plays an important
role in reducing the bias and resolution of vertex reconstruction.

The bias and resolution in Fig.~\ref{subfig:E_ratio_Q_fht} and Fig.~\ref{subfig:E_res_Q_fht}
are essentially the same as the results with charge alone, indicating
that the energy reconstruction is dominated
by the charge. The data points of energy resolution in
Figs.~\ref{subfig:E_res_Q} and \ref{subfig:E_res_Q_fht} are fitted with
Eq.~\eqref{eq:Eres_abc}\cite{abusleme:hal-04609915}:
\begin{equation}
  \frac{\sigma}{v(E)} =
  \sqrt{\left(\frac{a}{\sqrt{v(E)}}\right)^2
  + b^2 + \left(\frac{c}{v(E)}\right)^2},
  \label{eq:Eres_abc}
\end{equation}
where \(a\) denotes the Poisson statistical contribution from the PE
count; \(b\) is related to energy non-linearity and non-uniformity, including
quenching effect, Cherenkov radiation and electronics non-linearity; \(c\)
reflects the influence of dark noise. The best-fit results of \(a\),
\(b\) and \(c\) are \SI{2.002}{\%}, \SI{0.656}{\%} and \SI{8.31e-5}{\%},
respectively. The fitted energy resolution at \SI{1}{MeV} kinetic energy is \SI{2.07}{\%}.

\begin{figure*}[htbp]
  \centering
  \begin{subfigure}{0.42\textwidth}
    \centering
    \includegraphics[width=\textwidth]{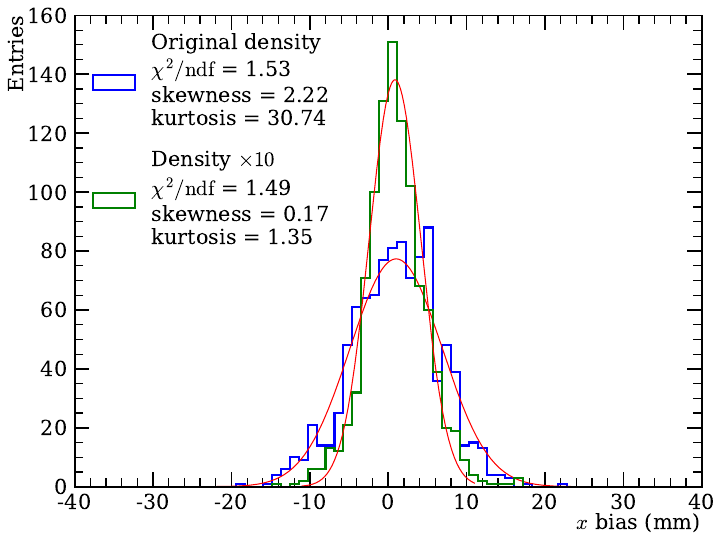}
    \caption{}
    \label{subfig:x_pos_distrib_0.5}
  \end{subfigure}
  \hspace{0.02\textwidth}
  \begin{subfigure}{0.42\textwidth}
    \centering
    \includegraphics[width=\textwidth]{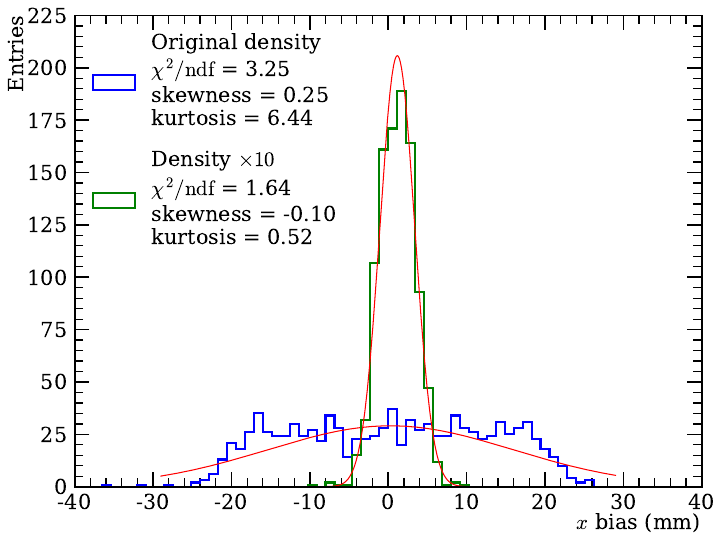}
    \caption{}
    \label{subfig:x_pos_distrib_5.0}
  \end{subfigure}
  \caption{Distribution of reconstructed \(x\) position for
  \SI{0.5}{MeV}~(\subref{subfig:x_pos_distrib_0.5}) and
  \SI{5.0}{MeV}~(\subref{subfig:x_pos_distrib_5.0}) electrons.
  Green and blue lines represents the
  two different GdLS densities. Red lines
  are fitted with Gaussian. At higher energy, the shape
  of distribution is affected by the electron tracks, breaking
  the point-like assumption.}
  \label{fig:x_pos_distrib}
\end{figure*}

The electron track effect is evident when we look closer into the
distribution of reconstructed vertices in
Fig.~\ref{fig:x_pos_distrib}. The flat-shaped vertex distribution of
electrons at \SI{5.0}{MeV} deviates from Gaussian. In our point-model,
a reconstructed vertex is the barycenter of the energy deposition
along the track, which is shifted from the starting point in the
direction of the \ce{e-} momentum.  The projection of isotropic shifts
onto the \(x\)-axis results in the flat distribution.  To verify our
speculation, we artificially enlarge the GdLS density 10 times
so that the mean free path of \SI{5.0}{MeV} \ce{e-} is less than
\SI{1}{mm}. The resulting distribution of reconstructed \(x\) shown in
green of Fig.~\ref{fig:x_pos_distrib} returns to Gaussian as expected.
For better modeling, we should no longer treat \ce{e-} of several
\si{MeV} as point sources.  With the introduction of tracks in our
future work, the reconstructed vertex should be the starting point of
the track rather than the energy-deposition barycenter.

Our tests support the feasibility of reconstruction using first hit time and charge
according to a pure probabilistic model introduced in
Sec.~\ref{sec:detector_response} and Sec.~\ref{subsec:charge_rec_lik}.
The high yield of scintillation photons leads to a Cherenkov photon
fraction of \(0.45\%\) in the first hit time, which contributes to a negligible
anisotropy in the model.

\subsection{With charge \(Q\) and smeared first hit time \(T_{\text{s}}\)}
In reality, the first hit time \(T\) is smeared by intrinsic transit
time spread~(TTS), but for SiPM TTS is at the level of \SI{100}{ps}.
Discrete sampling of analog-digital converter~(ADC) and time walk
effects impose larger time uncertainty than TTS at TAO, though they
could in principle be mitigated by clever firmware design. In this
section, we consider two extreme cases of time blurring, \SI{0.1}{ns}
for TTS alone and \SI{8}{ns} for the sampling interval of ADC.

\begin{figure*}[htbp]
  \centering
  \begin{subfigure}{0.32\textwidth}
    \centering
    \includegraphics[width=\textwidth]{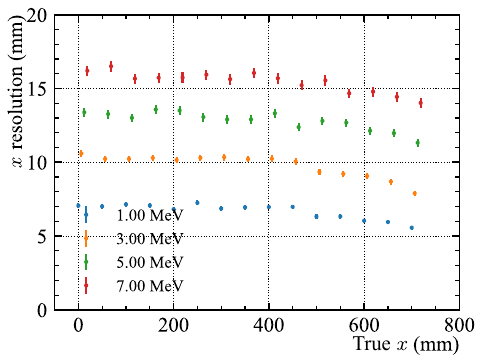}
    \caption{\SI{0.1}{ns} smear.}
    \label{subfig:x_res_Q_sfht_0.1}
  \end{subfigure}
  \begin{subfigure}{0.32\textwidth}
    \centering
    \includegraphics[width=\textwidth]{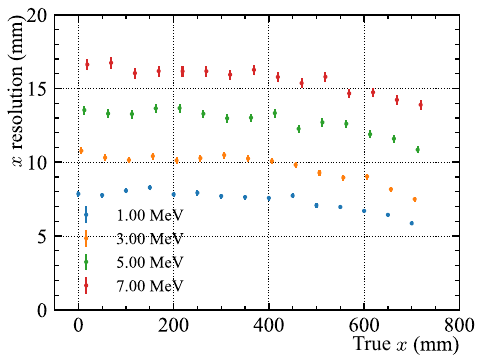}
    \caption{\SI{0.2}{ns} smear.}
    \label{subfig:x_res_Q_sfht_0.2}
  \end{subfigure}
  \begin{subfigure}{0.32\textwidth}
    \centering
    \includegraphics[width=\textwidth]{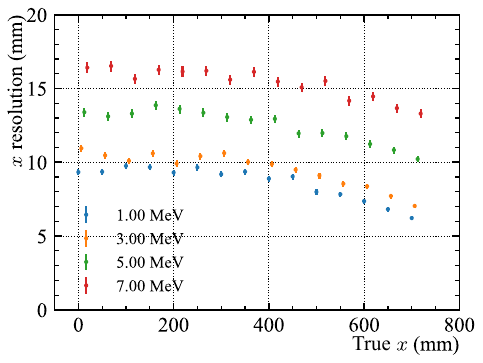}
    \caption{\SI{0.5}{ns} smear.}
    \label{subfig:x_res_Q_sfht_0.5}
  \end{subfigure}\\
  \begin{subfigure}{0.32\textwidth}
    \centering
    \includegraphics[width=\textwidth]{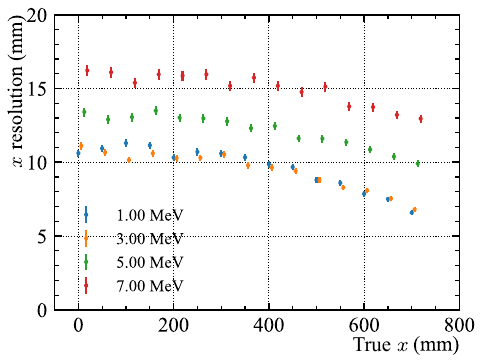}
    \caption{\SI{1}{ns} smear.}
    \label{subfig:x_res_Q_sfht_1.0}
  \end{subfigure}
  \begin{subfigure}{0.32\textwidth}
    \centering
    \includegraphics[width=\textwidth]{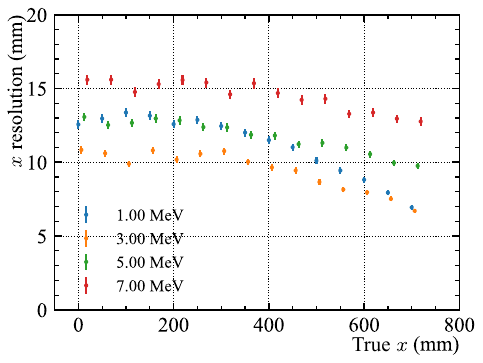}
    \caption{\SI{4}{ns} smear.}
    \label{subfig:x_res_Q_sfht_4.0}
  \end{subfigure}
  \begin{subfigure}{0.32\textwidth}
    \centering
    \includegraphics[width=\textwidth]{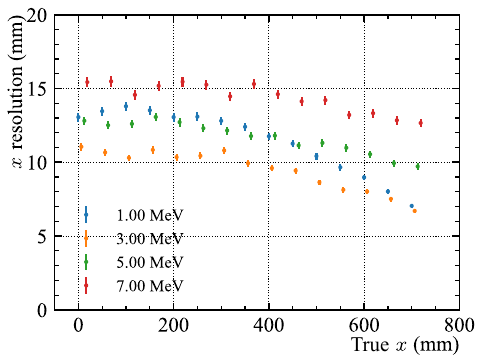}
    \caption{\SI{8}{ns} smear.}
    \label{subfig:x_res_Q_sfht_8.0}
  \end{subfigure}
  \caption{The reconstruction results of position resolution with charge \(Q\)
  and smeared first time \(T_{\text{s}}\). (\subref{subfig:x_res_Q_sfht_0.1})--(\subref{subfig:x_res_Q_sfht_8.0}) show the
  changes of vertex resolution for blurring \(\sigma = 0.1\) to \SI{8.0}{ns}.
  \SI{1}{MeV} and lower energy events are affected the most by time
  accuracy while \SI{5}{MeV} and higher energy events are dominated by
  their track effect.}
  \label{fig:t_sigma}
\end{figure*}

Without loss of generality, we add a Gaussian smear
\(\Delta T \sim N(0, \sigma^{2})\) to each PE in a SiPM channel, where
the first PE might be overtaken by the second one after the smearing.
The updated \emph{smeared first hit time} \(T_{\text{s}j}\)
substitutes \(T_j\) in Eq.~\eqref{eq:fht_charge_lik}. Meanwhile, the
response function Eq.~\eqref{eq:pmt_response_function} is convoluted
with the same Gaussian kernel. Fig.~\ref{fig:t_sigma} gives a series
of position resolution plots for \(\sigma\) running from \(0.1\) to
\SI{8.0}{ns}. Only position resolution of \SI{\sim 1}{MeV} \ce{e-}
changes significantly because that of higher energy \ce{e-} is
dominated by track effect~(Fig.~\ref{fig:x_pos_distrib}).  As
\(\sigma\) becomes larger, the resolution of \SI{\sim 1}{MeV}
transits from time to charge dominance.

Energy reconstruction with \(Q\) and \(T_{\text{s}}\) are consistent
with those in Fig.~\ref{fig:x_E_rec}.

\begin{figure*}[htbp]
  \centering
  \begin{subfigure}{0.4\textwidth}
    \centering
    \includegraphics[width=\textwidth]{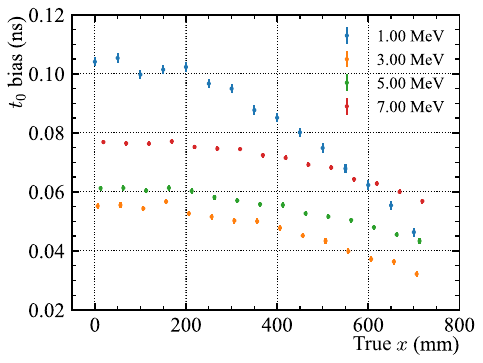}
    \caption{}
    \label{subfig:t_bias_Q_fht}
  \end{subfigure}
  \begin{subfigure}{0.4\textwidth}
    \centering
    \includegraphics[width=\textwidth]{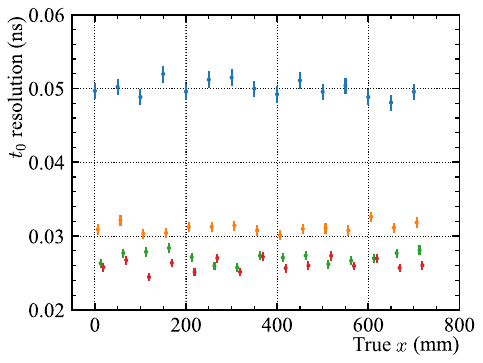}
    \caption{}
    \label{subfig:t_res_Q_fht}
  \end{subfigure}
  \caption{Bias (\subref{subfig:t_bias_Q_fht}) and resolution
    (\subref{subfig:t_res_Q_fht}) of the reconstructed event time \(t_0\) with
    \(1.0 \sigma\) smeared first hit time.}
  \label{fig:t_rec}
\end{figure*}

Because of time in likelihood function~Eq.~\eqref{eq:fht_charge_lik},
the algorithm can also give the reconstructed event time \(t_0\).
Figs.~\ref{subfig:t_bias_Q_fht} and \ref{subfig:t_res_Q_fht}
are the bias and resolution of time reconstruction respectively,
which shows the same trend over energy,
as the reconstruction results of vertex resolution using only charge.
The time bias is the difference between reconstructed event time
\(t_0\) and the real event time (default zero)
in detector simulation.
In reality \(t_0\) is affected by
trigger time and time delay in cable~\cite{ziyuan2021},
thus the result of time bias is provided as a reference.
The time resolution without considering TTS is less than
\SI{0.02}{ns}.


\section{Discussion}
\label{sec:discussion}

Our model is applicable to other neutrino detectors. There are several
points to be improved.

\subsection{Related work}

Z.~Li~et~al.~\cite{ziyuan2021} use the charge to estimate
the PE count roughly, according to average charge of one PE on PMT.
The construction of nPE map in G.~Huang's work~\cite{guihong2023}
shares the similar shortcoming. Due to the fluctuation of charge for one PE,
it is impossible to get an accurate PE count considering only one charge
value. Although waveform analysis~\cite{Xu_2022} is helpful to determine
the PE count and timing, it cannot be applied to time and
charge readouts. Tweedie distribution takes into account the fluctuations of PE count
and charge, thus inherently solving the above problem.
Nonetheless, the dependency between first hit time \(T\) and PE count \(N\)
is also important. It is the foundation to understand the time-charge
dependency and reconstruct with \(T\) and \(N\).
The joint distribution \(f_{\text{TN}}\left[T, N;R(t)\right]\) can be derived with
a similar method discussed in Sec.~\ref{subsec:TQ_distribution}.
Or just simply replace the Tweedie distribution
\(f_{\text{Tw}} (Q;\lambda)\)
in Eq.~\eqref{eq:TQ} with
the Poisson probability of \(N\)
\begin{equation}
   p_{\pi}(N;\lambda) = \exp(-\lambda)\lambda^N / N!,
\end{equation}
and then derive the joint distribution
\begin{equation}
   f_{\text{TN}}\left[T, N;R(t)\right] = \frac{\exp(-\lambda_{[\underline{T}, \overline{T}]})
   R(T) \lambda_{[T, \overline{T}]}^{N-1}
   }{(N - 1)!}.
   \label{eq:TN}
\end{equation}
The normalization of Eq.~\eqref{eq:TN} can also be verified.
Eq.~\eqref{eq:TN} is so-called first photoelectron timing technique~\cite{ziyuan2021},
which was also derived by G.~Ranucci~\cite{RANUCCI1995389}, later by C.~Galbiati and K.~McCarty~\cite{GALBIATI2006700}.
The form of reconstruction likelihood is similar to
Eq.~\eqref{eq:fht_charge_lik}:

\begin{equation}
\begin{aligned}
  &L\left( \vec{r}, v_{E}, t_0; \{(T_{j}, N_{j})\}\right)\\
  =& \prod_{\substack{N_{j}>0\\\text{hit}}} f_{\text{TN}}\left[T_j,N_j;v_{E}R_{j}^{0}(t
  - t_{0};\vec{r})\right]\\
   &\times \prod_{\substack{N_{j}=0\\ \text{nonhit}}}
  p_{\pi}\left(0;v_{E}\lambda^{0}_{j,[\underline{T} - t_0, \overline{T} -
  t_0]}\left(\vec{r}\right)\right) \\
  =& \prod_{N_{j} > 0} \biggl\{
    \frac{1}{(N_j - 1)!}
    \exp\left[-v_{E}\lambda^{0}_{j,[\underline{T} - t_0, \overline{T} -
    t_0]}(\vec{r})\right]\\
   &\quad v_{E} R^{0}_j\left(T_j - t_0;\vec{r}\right)
    \bigl[v_{E} \lambda^{0}_{j,[T_{j} - t_0, \overline{T} -
     t_0]}(\vec{r})\bigr]^{N_{j} - 1}\biggr\} \\
  &\times \prod_{N_{j} = 0} \exp\left[ -v_{E}\lambda^{0}_{j,[\underline{T} - t_0,
    \overline{T} - t_0]}(\vec{r})\right].
\end{aligned}
\label{eq:fht_npe_lik}
\end{equation}

\subsection{Application of Tweedie GLM on PMT}

\begin{figure}[htbp]
  \centering
  \includegraphics[width=0.4\textwidth]{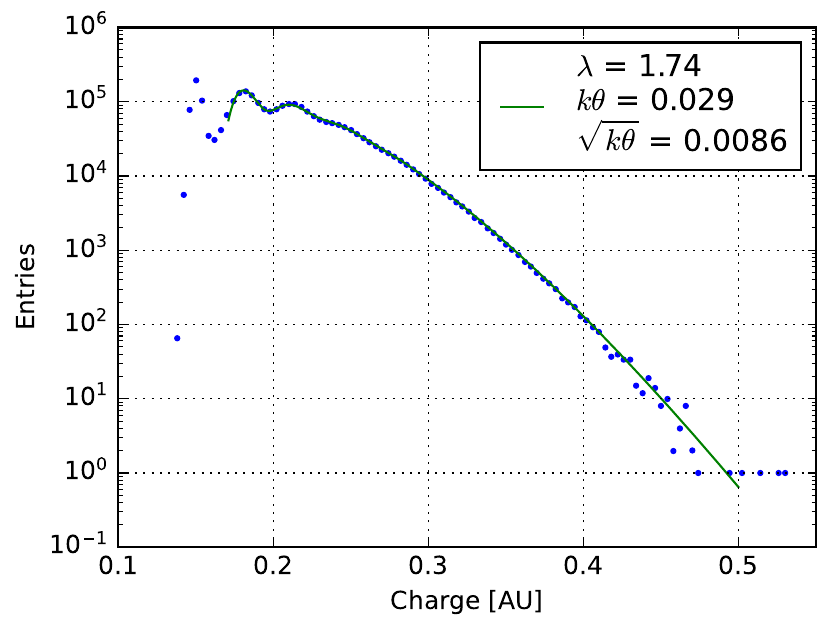}
  \caption{Charge distribution of PMT (blue point) and fitting result
    using Tweedie distribution (solid line). The first peak is the
    pedestal and the second one is the single photoelectron peak.
    The proportion of charge near the pedestal is about 0.175, which
    is consistent with the fitting results \(\exp\left(-\lambda\right)\approx0.176\).
    The charge spectrum is extracted from Fig.~1 in Kalousis's
    report~\cite{KALOUSIS2024168943}.}
  \label{fig:tweedie_pmt}
\end{figure}

Compared to SiPM's charge spectrum, that of PMT has greater variance
in Gamma part of Tweedie distribution, and it can also be fitted with
Tweedie GLM.  Fig.~\ref{fig:tweedie_pmt} shows a charge spectrum of
PMT extracted from Fig.~1 in Kalousis's
report~\cite{KALOUSIS2024168943} and the fitting result using Tweedie
distribution. The charges around the pedestal are neglected in the
fitting and considered as zero. It indicates that Tweedie GLM is not
only suitable for charge distribution of SiPM with low crosstalk,
but also for PMT spectrum modeled by
Kalousis~\cite{KALOUSIS2024168943} and
Anthony~et~al.~\cite{Anthony_2018}.

\subsection{The importance of \(T\) on other detectors}

Figs.~\ref{subfig:E_ratio_Q}, \ref{subfig:E_ratio_Q_fht}, \ref{subfig:E_res_Q}
and \ref{subfig:E_res_Q_fht} indicates that the time inputs have negligible
improvement on energy reconstruction for TAO. Due to high photo-coverage (\si{\sim94\%}) and photon detection efficiency (\si{ > 50\%}) of TAO detector,
charge-only point-like reconstruction is comparable to the T-Q combined one.
Both of them introduce little non-uniformity.

Nevertheless, the time inputs and the T-Q combined likelihood~(Eq.~\ref{eq:fht_charge_lik})  have significant potential for
vertex and energy reconstruction at larger LS detectors.
Position reconstruction is more sensitive to
time than charge, and it impacts on energy resolution via non-uniformity.

\subsection{Track effect in reconstruction}
\label{sec:track}

In Sec.~\ref{sec:dataset}, we achieve a vertex position resolution better than
\SI{20}{mm} for point-like events, which is much greater than the requirement in
TAO CDR~\cite{JUNO_2020ijm}. We find the non-Gaussian
distribution of position reconstruction in Fig~\ref{subfig:x_pos_distrib_0.5}
and worsening resolution of vertex position with \ce{e-}
energy in Figs.~\ref{subfig:x_res_Q} and \ref{subfig:x_res_Q_fht}. Both
imply that traditional assumption of point-like source is not
appropriate for \si{MeV} \ce{e-}.

The effect is more manifested with \ce{e+} and \(\gamma\) as their energy deposits are multi-sited.
It demands to extend the point-like model to
a track-like one for meticulous reconstruction.  The prerequisite is
precise time measurement in electronics, which is crucial to ameliorate position resolution~(Fig.~\ref{fig:t_sigma}).

The dynamics of track-like events are governed by the physics laws of positron
annihilation, Compton scattering and photoemission of electrons.  Embedding them
into reconstruction will give powerful constraints on the allowed parameter space.
It will lead to rigorous estimates of vertex and momentum of an incident particle.
Additionally, it is helpful to moderate the impact of energy leakage on
\ce{e+} and \(\gamma\) energy resolution.

Future track-like reconstruction depends strongly on the
precise time measurement. It is important to deploy ADC with
higher time precision and develop electronics firmware with
the advanced time-over-threshold~\cite{ota_dual_2019,wang_multiple_2020} to improve time resolution.

\subsection{Calibration of the optical detector model}
Our model is based on Monte Carlo simulation of \SI{0.5}{\MeV} \ce{e-}
to give a prefect point-source response.
However, the most common radioactive source deployed in detector
calibration is the \(\gamma\) source, such as \ce{^{137}Cs} and
\ce{^{60}Co}. The \(\gamma\) deposits energy at the scale of
\SI{10}{cm} and cannot be used directly to construct \(R^0(\cdot)\)
in Eq.~\ref{eq:pmt_response_function}.
We are developing a robust algorithm to extract
point-source response from \(\gamma\) calibrations by properly modeling
the track-effects of \(\gamma\).

\subsection{External crosstalk}
\emph{External crosstalk} or \emph{optical crosstalk} is the processes
involving photon emission of a SiPM that registers PEs on the surrounding SiPMs.
The mechanism can be compared to PMT flashers.
External crosstalk is suitable to be included in the optical detector model and
should not be included in the charge model,
because it involves multiple SiPMs and resembles diffuse reflections of photons.
We shall model the external crosstalk after the \emph{in situ} characterization
of external crosstalk of SiPM is obtained during TAO commissioning.


\section{Conclusion}
\label{sec:conclusion}

From first principles,
a pure probabilistic methodology is proposed
to simultaneously reconstruct vertex, energy and time
for point-like events in TAO CD and shown to meet the requirement and perform well.
In fiducial volume of TAO detector and energy range of reactor
neutrinos, after considering the dark noise and direct crosstalk
of SiPMs, for \SI{1}{\MeV} \ce{e-}, position resolution better than \SI{20}{mm} energy resolution of \SI{2}{\percent}
is achieved.  It does not impose extra non-linearity from reconstruction, controlling it within \SI{0.4}{\percent}.
Owing to high photon-detection efficiency and precise time measurement,
the track effect for \si{MeV} \(e^{-}\) is evident.
This methodology sufficiently utilizes
first hit time and charge
in reconstruction, which can be used not only for
SiPM in TAO detector, but also for other experiments with
first hit time and charge readouts, such as 3-inch PMT in
JUNO~\cite{Conforti_2021} and QBEE electronics in
Super-Kamiokande~\cite{5446533}.


\bmhead{Acknowledgements}

We are grateful to Yuyi~Wang for proofreading the manuscript.
We also thank Yiyang~Wu, Jun~Weng and Aiqiang~Zhang for
discussions of reconstruction algorithms
and development of SiPM/PMT charge model using Poisson and Gamma distributions.
We appreciate the help of JUNO collaboration which keeps our research on track.

The idea of first-principle time-charge likelihood originates from the draft
\emph{va3} fitter at KamLAND~\cite{xu_observation_2014}.  It grew out
of the reconstruction parameter tuning guided by Prof.~Itaru Shimizu,
and received scrutinization and encouragement from Prof.~Jason
Detwiler. The corresponding author would like to sincerely appreciate
the KamLAND collaboration for education and inspirations on event
reconstruction and neutrino physics.
This work is supported by the National Natural Science Foundation of China (No.123B2078).

\begin{appendices}

    \section{The normalization of \(f_{\text{TQ}}\)}
\label{appendix_A}

To verify the normalization of Eq.~\eqref{eq:TQ}, first integrate \(T\):
\begin{equation}
  {\small
    \begin{aligned}
        & \int_{\underline{T}}^{\overline{T}} f_{\text{TQ}}\left[T,Q;R(t)\right]
      \mathrm{d}T                                                                                                \\
      = & -\int_{\lambda_{[\underline{T}, \overline{T}]}}^{0} \exp\left[-\bigl(\lambda_{
          [\underline{T}, \overline{T}]} - \lambda\bigr)\right]
      \left(1 + \frac{\partial}{\partial \lambda}\right)f_{\text{Tw}} \left(Q ; \lambda \right)\mathrm{d}\lambda \\
      = & -\int_{\lambda_{[\underline{T},
              \overline{T}]}}^{0} \frac{\partial}{\partial \lambda}
      \left\{\exp\left[-\bigl(\lambda_{[\underline{T},
          \overline{T}]} - \lambda\bigr)\right]
      f_{\text{Tw}} \left(Q ; \lambda \right) \right\} \mathrm{d}\lambda                                         \\
      = & - \left.\exp\left[-\bigl(\lambda_{[\underline{T},
          \overline{T}]} - \lambda\bigr)\right]
      f_{\text{Tw}} \left(Q ; \lambda \right)
      \right|_{\lambda_{[\underline{T},
      \overline{T}]}}^{0}                                                                                        \\
      = & f_{\text{Tw}} (Q ; \lambda_{[\underline{T},
            \overline{T}]} ),
    \end{aligned}
  }
  \label{eq:TQ_int_tq}
\end{equation}
Eq.~\eqref{eq:TQ_int_tq} is the Tweedie PDF in \ref{subsec:charge_rec_lik}.
The first line of Eq.~\eqref{eq:TQ_int_tq} uses
\begin{equation}
  \mathrm{d} \lambda = - R(T) \mathrm{d}T.
\end{equation}
Then integrate \(Q\),
obviously
\begin{equation}
  \int f_{\text{Tw}}(Q ; \lambda_{[\underline{T},
        \overline{T}]}) \mathrm{d} Q = 1.
\end{equation}
Of course, we can first integrate \(Q\), but notice if \(Q\) is zero:
\begin{equation}
  f_{\text{Tw}} (Q = 0;\lambda_{[\underline{T},
        \overline{T}]}) = \exp(-\lambda_{[\underline{T},
    \overline{T}]}),
\end{equation}
\begin{equation}
  \begin{aligned}
      & f_{\text{T}}\left[T; Q \ne 0, R(t)\right]                               \\
    = & \int f_{\text{TQ}}\left[T, Q ; Q \ne 0, R(t)\right] \mathrm{d}Q         \\
    = & \frac{1}{1 - \exp(-\lambda_{[\underline{T}, \overline{T}]})}
    \biggl\{\int \exp\bigl[-\int_{\underline{T}}^{T} R(t) \mathrm{d}t\bigr]
    R(T)                                                                        \\
      & \times \left.\left(1 +
    \frac{\partial}{\partial \lambda}\right) f_{\text{Tw}} \left(Q ;
    \lambda \right)\right|_{\lambda = \int_{T}^{\overline{T}} R(t)
    \mathrm{d}t} \mathrm{d}Q \biggr\}                                           \\
    = & \frac{\exp\left[-\int_{\underline{T}}^{T} R(t) \mathrm{d}t\right] R(T)}
    {1 - \exp(-\lambda_{[\underline{T},
        \overline{T}]})}.
  \end{aligned}
  \label{eq:first_hit_time}
\end{equation}
Eq.~\eqref{eq:first_hit_time} is the distribution of first hit time \(T\).
Then integrate \(T\),
\begin{equation}
  \begin{aligned}
      & \int_{\underline{T}} ^{\overline{T}} f_{\text{T}}\left[T; Q \ne 0,
    R(t)\right] \mathrm{d}T                                                \\
    = & \frac{-\int_{\lambda_{[\underline{T},
              \overline{T}]}}^{0}\exp\left[-(\lambda_{[\underline{T},
              \overline{T}]} -
        \lambda)\right] \mathrm{d}\lambda}
    {1 - \exp( - \lambda_{[\underline{T},
        \overline{T}]})}\\
    = & 1.
  \end{aligned}
\end{equation}


\end{appendices}

\bibliography{sn-bibliography}

\end{document}